\let\LaTeXcline\cline\documentclass[pdflatex,sn-basic]{sn-jnl}\let\cline\LaTeXcline

\usepackage{array}
\usepackage{changepage}
\usepackage{graphicx}%
\usepackage{multirow}%

\usepackage{amsmath,amssymb,amsfonts}%
\usepackage{mathrsfs}%
\usepackage[title]{appendix}%
\usepackage[table,xcdraw]{xcolor}%
\usepackage{booktabs}%

\usepackage[utf8]{inputenc}
\usepackage{enumitem}
\usepackage{caption}
\usepackage{pdfpages}
\usepackage{longtable}
\usepackage{arydshln}
\usepackage{mathabx}
\usepackage{threeparttablex}

 \defcitealias{ISO25010}{ISO, 2011}

\begin{document}

\title[Article Title]{Quality attributes of test cases and test suites -- importance \& challenges from practitioners' perspectives}

\author*[1]{\fnm{Huynh Khanh Vi} \sur{Tran}}\email{huynh.khanh.vi.tran@bth.se}

\author[1]{\fnm{ Nauman bin} \sur{Ali}}\email{nauman.ali@bth.se}

\author[1]{\fnm{Michael} \sur{Unterkalmsteiner}}\email{michael.unterkalmsteiner@bth.se}

\author[1]{\fnm{J\"urgen} \sur{B\"orstler}}\email{jurgen.borstler@bth.se}

\author[2]{\fnm{Panagiota} \sur{Chatzipetrou}}\email{panagiota.chatzipetrou@oru.se}

\affil*[1]{\orgdiv{Department of Software Engineering}, \orgname{Blekinge Institute of Technology}, \orgaddress{\street{Valhallav\"agen 1}, \city{Karlskrona}, \postcode{371 41}, \state{Blekinge}, \country{Sweden}}}

\affil[2]{\orgdiv{Department of Informatics}, \orgname{Centre for Empirical Research on Information Systems (CERIS), School of Business, \"Orebro University}, \orgaddress{\street{Fakultetsgatan 1}, \city{\"Orebro}, \postcode{701 82}, \state{\"Orebro}, \country{Sweden}}}

\abstract{
\textbf{Context:} The quality of the test suites and the constituent test cases significantly impacts confidence in software testing. While research has identified several quality attributes of test cases and test suites, there is a need for a better understanding of their relative importance in practice.
\textbf{Objective:} We investigate practitioners' perceptions regarding the relative importance of quality attributes of test cases and test suites and the challenges that they face in ensuring the perceived important quality attributes.
\textbf{Method:} To capture the practitioners' perceptions, we conducted an industrial survey using a questionnaire based on the quality attributes identified in an extensive literature review.
We used a sampling strategy that leverages LinkedIn to draw a large and heterogeneous sample of professionals with experience in software testing.
\textbf{Results:} We collected 354 responses from practitioners with a wide range of experience (from less than one year to 42 years of experience).
We found that the majority of practitioners rated Fault Detection, Usability, Maintainability, Reliability, and Coverage to be the most important quality attributes.
Resource Efficiency, Reusability, and Simplicity received the most divergent opinions, which, according to our analysis, depend on the software-testing contexts.
Also, we identified common challenges that apply to the important attributes, namely inadequate definition, lack of useful metrics, lack of an established review process, and lack of external support.
\textbf{Conclusion:} The findings point out where practitioners actually need further support with respect to achieving high-quality test cases and test suites under different software testing contexts.
Hence, the findings can serve as a guideline for academic researchers when looking for research directions on the topic.
Furthermore, the findings can be used to encourage companies to provide more support to practitioners to achieve high-quality test cases and test suites.
}


\maketitle

\section{Introduction} \label{sec:introduction}
There have been several studies over the years emphasizing the value of high-quality test cases and test suites~\citep{athanasiou2014test,grano2018empirical,garousi2018smells} as well as the effects of design flaws in them~\citep{spadini2018relation,kovacs_test_2014,garousi2018smells,garousi_what_2019}.
Moreover, modern software development with continuous software engineering~\citep{fitzgerald2017continuous} and DevOps~\citep{jabbari2018towards} relies extensively on automated and continuous testing.
For these reasons, it is important to ensure the quality of test cases and test suites.

To characterize and assess the quality of test cases and test suites, attributes similar to software product quality attributes have been proposed.
The most recent systematic literature review on the quality attributes of test cases and test suites~\citep{tran2021assessing} was conducted by the authors of this study.
This review provided a comprehensive overview of 30 quality attributes for test cases and test suites, and can serve as a guideline for both practitioners and researchers when searching for literature on the topic.
For high-quality test cases and test suites, one can argue that all 30 quality attributes should be relevant.
However, focusing on all quality attributes together can be very challenging, especially when certain trade-offs are expected between the quality attributes, i.e., all the attributes cannot be improved simultaneously.
The existence of trade-offs is well motivated by several studies that investigated software quality attributes~\citep{sas2020Quality,wahler2012quality,feitosa2015investigating, ali2012testing}.

Instead of focusing on all quality attributes or some random attributes, we argue that it is better to understand which quality attributes practitioners perceive as important so that we, as researchers, can focus on providing further support in assessing, maintaining, and improving these important attributes.
To provide such assistance to practitioners, there is a need to understand the software-testing contexts in which certain quality attributes are perceived as more or less important, as well as any challenges that practitioners face in defining, measuring, and maintaining these quality attributes.
Note that the referred contexts can be characterized by different context dimensions such as testing level, testing practice, and characteristics of the system under test~\citep{tran2021assessing}.

The quality of test cases and test suites has received great attention from the research community for several decades.
There also exist studies focusing on challenges in software testing in general~\citep{bertolino2007software,garousi2017industry,garousi2020exploring,santos2019mind,fulcini2023review,Muhammad2021design, ali2012testing}, or challenges related to the quality of test cases and test suites~\citep{juhnke2021challenges,kochhar2019practitioners}. 
However, to the best of our knowledge, there are no studies investigating practitioners' perceptions of the importance and challenges (with respect to defining, measuring, and maintaining) of a wide range of quality attributes of test cases and test suites in multiple contexts.
We conducted an industrial survey to address this research gap.
Our survey recruited practitioners from 20 LinkedIn groups to draw a large-scale sample and increase subject heterogeneity.
As a result, we received responses from 354 practitioners who got involved in various testing activities with a wide range of working experience.

The contribution of this survey study is three-fold: (1) the identification of generally important (i.e., independent of context) quality attributes of test cases and test suites in practice, namely Fault Detection, Maintainability, Reliability, Usability, and Coverage; (2) the confirmation of the influence of software-testing context dimensions (Automation activity, Testing activity, Testing level, Testing practice, Type of development process of System under test (SUT), Application domain of SUT, and Type of SUT) on the perceived importance of certain attributes, namely Resource Efficiency, Reusability, Simplicity, and Usability; and (3) the identification of common challenges with respect to defining, measuring, and maintaining the important attributes: inadequate definition, lack of useful metrics, lack of an established review process, and lack of external support.

Ultimately, our study provides valuable insights into how practitioners perceive the importance of quality attributes of both test cases and test suites.
Our findings highlight important quality attributes in practice, the weight of certain context dimensions on the importance of several quality attributes, and the challenges that apply to the important attributes.
Altogether, our study points out where practitioners actually need further support with respect to achieving high-quality test cases and test suites.
Hence, our findings can serve as a guideline for academic researchers when looking for research directions on the topic, knowing that their research can meet the actual needs in practice. 
Additionally, the findings can be used to encourage companies to provide more support to practitioners to achieve high-quality test cases and test suites.
Ultimately, our findings provide extra support for the knowledge transfer between academia and industry.

The rest of the paper is structured as follows: Section~\ref{sec:relatedWork} presents related work on the quality of test cases and test suites, general challenges in software testing, and challenges related to the quality of test cases and test suites.
Section~\ref{sec:researchMethod} describes how we conducted the survey.
Section~\ref{sec:results_analysis} presents the survey's results and our analysis for each research question.
In Section~\ref{sec:discussion}, we discuss selected findings related to the importance of quality attributes, their correlation to the software-testing contexts, and the challenges that apply to the important quality attributes.
Section~\ref{sec:validityThreats} discusses the limitations and the primary threats to validity.
Finally, Section~\ref{sec:conclusion} concludes our paper.

\section{Related work} \label{sec:relatedWork}
In this section, we discuss related work from three angles.
First, we look at past research on the quality of test cases and test suites.
Then, we discuss related works that address general challenges in software testing~\citep{bertolino2007software,garousi2017industry,garousi2020exploring,santos2019mind,fulcini2023review,Muhammad2021design, ali2012testing}.
Finally, we discuss studies that focus on challenges related to the quality of test cases and test suites~\citep{kochhar2019practitioners, juhnke2021challenges}.

\subsection{Quality of test cases and test suites}\label{sec:relatedWork_TC/TSQuality}
The quality of test cases and test suites has been investigated for several decades.
One of the earliest related works was conducted by Goodenough and Gerhard~\citeyearpar{goodenough1975toward}, who received attention from the research community when studying test adequacy and proposing a definition for reliable test data.
Many years later, Zhu et al.~\citeyearpar{zhu1997software} conducted a literature review on 
test adequacy criteria.
Their reviews encouraged fellow researchers to investigate the topic of structural test quality for specific application domains and programming paradigms~\citep{kapfhammer2003family,lemos2007control,pei2019deepxplore}.
Besides, some researchers turned their focus to adapting software quality models to define the quality of test cases and test suites~\citep{neukirchen2008approach,athanasiou2014test}.
In recent years, researchers started looking into how inputs from practitioners could help to construct useful quality models for test cases and test suites~\citep{bowes2017how,grano2020pinsa,tran2019test}.

The most recent literature review provides a consolidated overview of state of the art in test artifact quality in software testing~\citep{tran2021assessing}, conducted by the authors of this study.
The main contribution is a quality model of test cases and test suites containing 30 quality attributes, eleven of which have measurement information.
The quality model also includes quality attributes adapted from ISO/IEC 25010:2011 to provide a more extensive overview of the quality of test cases and test suites. 
In addition, we identified in the literature review eleven context dimensions that have been used to describe the software-testing contexts in which the quality has been studied. 
The quality model could assist in developing guidelines or templates for designing test cases and test suites and assessment tools for existing test cases and test suites.

The quality of test cases and test suites have also been discussed in studies on test smells.
The concept of test smells was coined by Van Deursen et al.~\citeyearpar{van2001refactoring}.
In the context of unit testing for eXtreme Programming, Van Deursen et al. define test smells as derived from code smells and as indicators of troubles in test code.
Later on, Meszaros~\citeyearpar{meszaros2007xunit}, while referring to the work of Van Deursen et al., introduced definitions of more test smells for the unit testing framework xUnit.
The latest secondary study on definitions of test smells was conducted by Garousi et al.~\citeyearpar{garousi2018smells}.
The study provided a list of 139 test smells, as well as a summary of approaches and tools to deal with the test smells.
Further in this direction, Aljedaani et al.~\citeyearpar{aljedaani2021test} provided an overview of the characteristics of over 22 test smell detection tools developed in the research community.

\subsection{General challenges in software testing}\label{sec:relatedWork_generalChallenges}
Bertolino~\citeyearpar{bertolino2007software} provided one of the first broad roadmaps of achievements and challenges to address in the field of software testing research. 
Particularly, the author highlighted certain important milestones, including matured research in (1) techniques and tools to support test design; (2) test criteria and their effectiveness; (3) different testing techniques.
The author emphasized the importance of a universal test theory, i.e., a coherent and rigorous framework that practitioners can refer to for selecting adequate testing approaches.
Besides, model-based testing, 100\% automatic testing, and cost-effective test engineering were also mentioned as what the future software-testing research should focus on.
For each of these foci, several challenges were highlighted.
For example, \textit{test effectiveness} (How effective is a test selection criterion for finding faults?) was listed as one of the main challenges for achieving a universal test theory.
For achieving 100\% automatic testing, the research in test input generation led to advancements in theory but has limited industrial impact.
Likewise, understanding testing costs was reported as a challenge for ``efficacy-maximized test engineering''.

Ten years later, Garousi et al.~\citeyearpar{garousi2017industry,garousi2020exploring} presented 105 practitioners' opinions on challenges in different testing activities, including test-case design (criteria-based or based on human expertise), test scripting, test execution, test evaluation, test-result reporting, test management, test automation/tools, and others.
The authors asked practitioners to rank the level of challenges (from no challenge at all to lots of challenges) in each testing activity, and then to propose concrete topics they wanted researchers to focus on.
According to their findings, test management and automation/tools and
``other'' test activities were considered the most challenging testing activities.
Their main conclusion was that industry and academic focus areas were not aligned.
While researchers tended to focus on theoretical challenges such as search-based test-case design, practitioners wanted to improve the effectiveness and efficiency of software testing.
Going in the same direction as Garousi et al., Santos et al.~\citeyearpar{santos2019mind} conducted a mapping study, a quantitative study, and a focus group to analyze the misalignment regarding important aspects for improvement between the software-testing research community and practitioners.
Their main conclusion was that even though practitioners and researchers were both interested in test automation, practitioners would like to focus on identifying, understanding, and modifying existing testing tools and strategies while researchers paid more attention to developing new ones.

Some studies focus on testing challenges for specific types of system under tests (SUT), application domains, or testing approaches.
Fulcini et al.~\citeyearpar{fulcini2023review} conducted a multivocal literature review that assessed techniques, tools, and challenges related to gamified software testing.
According to the authors, gamification was a promising testing approach but was not yet well-established, and several challenges needed to be addressed.
The authors categorized the challenges into three groups: design improvements (related to adding new gamified mechanics or improving existing ones regarding several aspects such as ethical concerns or tool simplification), implementation improvements (related to the technical implementation of the gamified tool and frameworks such as tool reusability and scalability), and evaluation (related to evaluating gamified mechanics such as the needs for automatic analysis or expert evaluation).
The study of Waseem et al.~\citeyearpar{Muhammad2021design} was based on 106 survey responses and six interviews and focuses on microservice systems.
Besides findings regarding designing and monitoring such systems, the authors reported that system complexity was one of the main reasons making the testing activities (creating and implementing manual tests, integration testing) challenging and there were no dedicated solutions.
Different from these aforementioned studies~\citep{fulcini2023review,Muhammad2021design}, which broadly looked into either general aspects of software testing or challenges related to specific software-testing contexts, we focused on the quality attributes of test cases and test suites and their importance in relation to different context dimensions (testing level, testing practice, type of SUT, etc.).

\subsection{Challenges related to the quality of test cases and test suites}\label{sec:relatedWork_challenges_testArtifactQuality}
The third group of related work includes studies~\citep{juhnke2021challenges,kochhar2019practitioners} that are closer to our work as they also discussed the quality of test artifacts.
Juhnke et al.~\citeyearpar{juhnke2021challenges} conducted 17 interviews and surveyed 36 practitioners to collect and analyze challenges regarding test case specifications in the automotive domain.
They classified the challenges into nine categories: availability problems with input artifacts; content-related problems with input artifacts, and knowledge-related problems; lack of knowledge; the test case description; the test case specification content; processes; communication; quality assurance; and tools.
Their study showed that the latter four challenges were assessed as more frequently occurring and more critical by practitioners.

Kochhar et al.~\citeyearpar{kochhar2019practitioners} invited 282 practitioners to validate 29 hypotheses regarding characteristics of good test cases and testing practices.
The hypotheses covered six aspects: test case contents, size and complexity, coverage, maintainability, and bug detection.
Each hypothesis was ranked on a Likert scale from ``strongly agree'' to ``strongly disagree''.
The authors concluded that some hypotheses received quite controversial opinions, and hence, researchers were suggested to conduct deep empirical studies to analyze them.
As for practitioners, their hypotheses described characteristics of good test cases, which the authors claimed could be used as guidelines for novices to design high-quality test cases.

\subsection{Research gap}\label{sec:contribution}
We found that general challenges in software testing have been reported in the literature (Section~\ref{sec:relatedWork_generalChallenges}).
Another aspect related to the quality of test cases and test suites is test smells, which have been intensively investigated in recent years (Section~\ref{sec:relatedWork_TC/TSQuality}).
We identified two studies from Juhnke et al.~\citeyearpar{juhnke2021challenges} and Kochhar et al.~\citeyearpar{kochhar2019practitioners} that investigated challenges related to the quality of test cases and test suites from the practitioners' perspective (Section~\ref{sec:relatedWork_challenges_testArtifactQuality}). 

While Juhnke et al.~\citeyearpar{juhnke2021challenges} proposed nine categories of challenges related to test case specifications, their context is limited to the automotive domain.
Our study is not restricted to a single software-testing context.
We referred to our tertiary study~\citep{tran2021assessing} on the quality of test cases and test suites together with various taxonomies~\citep{forward2008taxonomy,kotonya2003towards,vijayasarathy2016choice,souza2017roost,ray2014large,atecsougullari2020automation} and the ISO/IEC/IEEE 29119-1 standard~\citeyearpar{ISO29119} for different context dimensions in order to capture different software-testing contexts in practice.
Kochhar et al.~\citeyearpar{kochhar2019practitioners} studied practitioners' opinions on different hypotheses on good test cases and testing practices.
While their study covered some quality attributes, their attributes set was smaller than ours.
Also, it was not clear in their paper how the attributes were chosen.
In contrast, our quality attributes were extracted from our tertiary study~\citep{tran2021assessing} on the quality of test cases and test suites.

Therefore, even though our study also focused on understanding practitioners' perceptions of the quality of test cases and test suites, we systematically studied a wider range of software-testing contexts and a larger number of quality attributes of both test cases and test suites.
On top of that, we have not found any study that has investigated the correlation between the software-testing context dimensions and the perceived importance of quality attributes.
We filled in this research gap, by means of this survey study.

\section{Research method}\label{sec:researchMethod}
\subsection{Research questions} \label{sec:RQ}
The goal of our survey was to understand which quality attributes of test cases and test suites are considered important and the challenges in defining, measuring, and maintaining the important attributes from the practitioners' perspectives.
To achieve the stated goal, we answered the following research questions.
\begin{itemize}
    \item RQ1. What is the relative importance of quality attributes of test cases and test suites from practitioners' perspectives?
    
    We wanted to understand if some quality attributes might be perceived as more important than others by practitioners.
    To answer this research question, we asked respondents to rank the importance (as three levels: \textit{Important} / \textit{Optional} / \textit{Not relevant at all}) of the quality attributes of either test cases or test suites.
    We distinguished between test cases and test suites in the ranking task because the quality attributes of test cases and test suites can share the same names but be defined differently.
    Furthermore, a test suite can contain multiple test cases, and hence, the perceived importance of test-suite quality attributes can be weighted differently than of test-case quality attributes.
    
    \item RQ2. To what extent does the software-testing context influence the perceived importance of the quality attributes of test cases and test suites?
    
    The software-testing context contains the following context dimensions: Testing activity, Testing level, Testing type, Testing practice, Automation activity, Application domain of system under test (SUT), Type of SUT, and Type of development process.
    We wanted to investigate if some quality attributes might be considered important in some contexts but not in different contexts.
    To answer this research question, we asked respondents to describe their software-testing context dimensions and then we used statistical tests to analyze the correlation between the context dimensions and the importance level of the quality attributes (RQ1's answer).
    
    \item RQ3. What are the challenges do practitioners face in ensuring the important quality attributes of test cases and test suites?
     
    To help practitioners achieve high-quality test cases and test suites with respect to the important quality attributes, we wanted to understand the challenges that practitioners face in defining, measuring, and maintaining these attributes.    
    To answer this research question, for each quality attribute, we asked respondents to select among three provided challenges~\citep{juhnke2021challenges}: Challenge $C_{def}$ - Inadequate definition, Challenge $C_{metric}$ - Lack of useful metrics, and Challenge $C_{review}$ - Lack of an established review process.
    We also offered a free-text field to collect other challenges. 
\end{itemize}

\subsection{Survey design}\label{sec:surveyDesign}
\begin{figure}
\begin{center}
\includegraphics[width=1\linewidth]{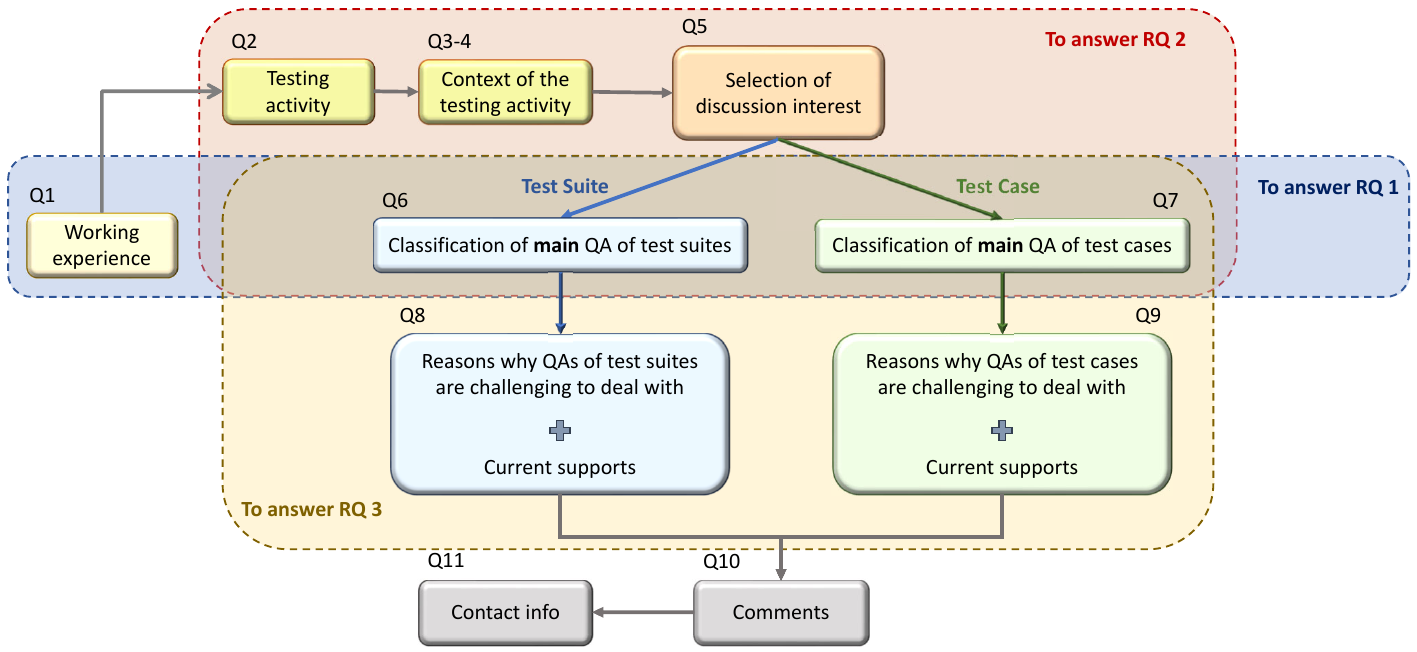}
\caption{Overview of the survey questionnaire}
\label{fig:surveyQuestionnaire}
\end{center}
\end{figure}

We followed Kitchenham and Pfleeger's guidelines~\citeyearpar{kitchenham2008personal} to design our survey, a structured cross-sectional web-based survey.
This approach helps to reach practitioners from geographically diverse locations and supports automated data collection.
The survey was administered via the web-based survey software, Qualtrics Survey\footnote{\url{https://www.qualtrics.com/uk/core-xm/survey-software/}}.
An overview of the questionnaire is shown in Figure~\ref{fig:surveyQuestionnaire}.

The questionnaire contains three sections.
The first section (Q1--4) collects demographic information about the practitioners (years of experience and software-testing contexts).
The second section (Q5--7) collects the perceived importance levels (\textit{Important} / \textit{Optional} / \textit{Not relevant at all}) of the quality attributes of test cases and test suites.
It is worth noting that we referred to several taxonomies~\citep{forward2008taxonomy,kotonya2003towards,vijayasarathy2016choice,souza2017roost,ray2014large,atecsougullari2020automation}, the ISO/IEC/IEEE 29119-1 standard~\citeyearpar{ISO29119} as well as our tertiary study~\citep{tran2021assessing} on the quality of test cases and test suites to provide the software-testing contexts and the list of quality attributes.
More details are presented further below in this section.

The third section (Q8--9) collects two sets of information: (i) challenges that apply to the \textit{important} quality attributes; (ii) information regarding any support the practitioners currently have to tackle the challenges, followed by the closing section (Q10--11).
The final version of the questionnaire can be viewed via \url{https://doi.org/10.6084/m9.figshare.23309702.v1}.

In the first section (Q1--4) of the questionnaire, we collected information regarding participants' working experience (Q1) and their software testing-related contexts (Q2--4) in which they have considered the quality of test cases and test suites to be relevant and challenging to achieve, assess, maintain, or improve.

To characterize the software-testing contexts (Q2--4), we first presented a list of categories of common software-testing activities, which were extracted from the ISO/IEC/IEEE 29119-1 standard~\citeyearpar{ISO29119}.
We then asked each participant to select a category that best describes his or her software-testing activities in which the participant has found it challenging to achieve, assess, maintain, or improve the quality of test cases or test suites (Q2).
The subsequent questions (Q3--4) were to gain a better understanding of the context of the selected testing activities.
For this purpose, we presented a list of context dimensions, each with common context values from which the participants could select.
For example, under the \textit{testing level} context dimension, a participant could select one of the following context values: unit testing, integration testing, system testing, system integration testing, and acceptance testing.

More specifically, we focused on the following context dimensions: testing level, testing type, testing practice, automation activity, and test suite's size, as well as the dimensions related to the systems under test (SUT): size, application domain, software type, and type of development process.
These context dimensions were aggregated in our tertiary study~\citep{tran2021assessing} on the quality of test cases and test suites.
Additionally, we relied on different taxonomies~\citep{forward2008taxonomy,kotonya2003towards,vijayasarathy2016choice,souza2017roost,ray2014large,atecsougullari2020automation} and the ISO/IEC/IEEE 29119-1 standard~\citeyearpar{ISO29119} to provide common context values under each context dimension in Q3--4.
It is worth noting that there were also free text fields where the respondents could enter their own category of testing activities (Q2) and context value (Q3--4) if they found none of the provided categories and values were appropriate.

To help us answer RQ2 (investigating the extent software-testing contexts influence the perceived importance of the quality attributes of test cases and test suites), we stated explicitly in Q3--4 that the respondent should ``consider the single context where you found the quality of test cases and test suites one of your main concerns.''
Due to this reason, we restricted each respondent to select only one context value under each context dimension.
Without this design approach, we might have received an answer in which, for example, the respondent provided a mixed context of unit testing and system testing and marked two quality attributes, \textit{Maintainability} and \textit{Fault Detection}, as \textit{Important}.
In this example, it would be impossible to tell which quality attribute was associated with which testing level.

The second section (Q5--7) of the questionnaire collected the backbone data of the study.
Q5 asked the respondents to select whether they wanted to share their opinions on the quality of either test cases or test suites.
The reason for having Q5 is explained under RQ1 in Section~\ref{sec:RQ}.
Depending on their answers to Q5, we asked the respondents to rank the importance of each main quality attribute of test cases (Q6) or test suites (Q7) as \textit{Important} / \textit{Optional} / \textit{Not relevant at all}.
Note that in Q6 and Q7, we reminded each respondent to think about the importance of each quality attribute in the software-testing context that he or she specified in Q2--4.
To facilitate this ranking task, we provided the respondents with a test-case quality model (to answer Q6) and a test-suite quality model (to answer Q7).
Each quality model consists of a list of main quality attributes and their sub-quality attributes, similar to the software quality models in ISO/IEC 25010:2011~\citeyearpar{ISO25010}.
Additionally, each quality attribute comes with a short description.
The quality models and the descriptions of all provided quality attributes were aggregated from our tertiary study~\citep{tran2021assessing} and the related work as shown in Table~\ref{tab:Quality_Attributes_Descriptions}.
Note that further quality attributes were presented in our tertiary study.
However, some of them did not have descriptions as they were not described in the selected secondary studies or in the related work.
Hence, these attributes were excluded from this survey.

{
    \begin{ThreePartTable}
    \begin{TableNotes}
    \footnotesize
    \noindent
    \begin{minipage}[c]{0.9\textwidth}
    \item \textbf{[*] Main quality attribute (QA) }
    \item \textbf{[$\drsh$] Sub-quality attribute}. Note that not all main QAs have a sub-QA(s).    
    \end{minipage} 
    \begin{minipage}[c]{0.3\textwidth}
    \item S1: \citep{tran2021assessing}
    \item S2: \citep{zeiss2007applying}
    \item S3: \citep{deOliveira2018visualizing}
    \item S4: \citep{chernak2001validating}
    \end{minipage} 
    \begin{minipage}[c]{0.3\textwidth}
    \item S5: \citep{adlemo2018test}
    \item S6: \citepalias{ISO25010}
    \item S7: \citep{garousi2015software}
    \item S8: \citep{zaidman2008mining}
    \end{minipage}
    \begin{minipage}[c]{0.3\textwidth}
    \item S9: \citep{tran2019test}
    \item S10: \citep{bowes2017how}
    \item S11: \citep{kochhar2019practitioners}
    \item S12: \citep{athanasiou2014test}
    \item S13: \citep{daka2015modeling}
    \end{minipage}
    \end{TableNotes}
    \footnotesize
    \setlength\dashlinedash{0.3pt}
    \setlength\dashlinegap{1.5pt}
    \setlength\arrayrulewidth{0.3pt}
    \begin{longtable}{p{0.17\textwidth}p{0.07\textwidth}p{0.35\textwidth}p{0.25\textwidth}p{0.035\textwidth}}
        \caption{Descriptions of main quality attributes and their sub-quality attributes of test suites (TS) and test cases (TC)}
        \label{tab:Quality_Attributes_Descriptions} \\ \hline
        \rowcolor[HTML]{D5D4D4} 
        \textbf{Quality Attribute} & \textbf{For} & \textbf{Description in case of test suites} & \textbf{Description in case of test cases} & \textbf{Ref} \\ \hline
        \endfirsthead
        \multicolumn{5}{c}%
        {{ Table \thetable\ continued from previous page}} \\ \\ \hline
        \rowcolor[HTML]{D5D4D4} 
        \textbf{Quality Attribute} & \textbf{For} & \textbf{Description in case of test suites} & \textbf{Description in case of test cases} & \textbf{Ref} \\ \hline
        \endhead
        \bottomrule
        \insertTableNotes 
        \endfoot

        \insertTableNotes 
        \endlastfoot
        
        \rowcolor[HTML]{ECF4FF}
        \textbf{Coverage} \tnote{*} & TS & The degree of completeness and effectiveness with which a test suite ensures that SUT code is executed and the system requirements are verified. & N/A & S1 \\ \hdashline
        $\drsh$ Code coverage completeness & TS & The degree to which a test suite ensures that the SUT code is executed. & N/A & S1 \\ \hdashline
        $\drsh$ Requirement/ feature coverage completeness & TS & The degree to which a test suite ensures that the system requirements are addressed. & N/A & S2 \\
        \hline
        
        \rowcolor[HTML]{ECF4FF}
        \textbf{Diversity} \tnote{*} & TS & The degree of dissimilarity among test cases in terms of static (e.g., test data, test steps, execution traces, code statements) or dynamic (e.g., execution history) information. & N/A & S1, S3 \\ 
        \hline
        
        \rowcolor[HTML]{ECF4FF}
        \textbf{Fault detection} \tnote{*} & TS, TC & \multicolumn{2}{p{0.632\textwidth}}{The degree of effectiveness and efficiency of identifying defects that a [test suite $\mid$ test case] can achieve.} & S1 \\ \hdashline
        $\drsh$ Fault detection capability/ effectiveness & TS, TC & \multicolumn{2}{p{0.632\textwidth}}{The more defects a [test suite $\mid$ test case] finds, the more effective it is.} & S1, S4 \\ \hdashline
        $\drsh$ Fault detection efficiency & TS, TC & \multicolumn{2}{p{0.632\textwidth}}{How quickly a [test suite $\mid$ test case] can discover faults.} & S1, S2 \\ 
        \hline
        
        \rowcolor[HTML]{ECF4FF}
        \textbf{Maintainability} \tnote{*} & TS, TC & \multicolumn{2}{p{0.632\textwidth}}{The capability of a [test suite $\mid$ test case] to be modified for error correction, improvement, or adaption to changes in the environment or the requirements.} & S5, S2 \\ 
        \hdashline
        $\drsh$ Analyzability & TS, TC & \multicolumn{2}{p{0.632\textwidth}}{The property of a [test suite $\mid$ test case] to be diagnosed for deficiencies or causes of failures in the [test suite $\mid$ test case], or for identifying which parts to be modified.} & S6, S2 \\ \hdashline
        $\drsh$ Co-evolution \& maintenance & TS, TC & \multicolumn{2}{p{0.632\textwidth}}{[Test suites $\mid$ Test cases]  and production code should be developed and maintained synchronously} & S7, S8 \\ \hdashline
        $\drsh$ Homogeneity & TS, TC & The degree to which the design of test cases in a test suite follows the same rules. & The degree to which the template for test cases follows the same rules. & S9 \\ \hdashline
        $\drsh$ Self-contained & TS, TC & The degree to which test cases in a test suite can be executed without relying on the execution of other test cases. & The degree to which a test case can be executed without relying on the execution of other test cases. & S1, S10, S5 \\ \hdashline
        $\drsh$ Traceability & TS, TC & The degree to which test cases in a test suite can be linked to other related artifacts such as tested requirements, features, source code, and issues. & The degree to which a test case can be linked to other related artifacts such as tested requirements, features, source code, and issues. & S1, S9, S5 \\ 
        \hline
        
        \rowcolor[HTML]{ECF4FF}
        \textbf{Reliability} \tnote{*} & TS, TC & \multicolumn{2}{p{0.632\textwidth}}{The capability of a [test suite $\mid$ test case] to maintain a specific level of performance under different conditions. ``Performance'' expresses the degree to which specific needs and requirements towards the [test suite $\mid$ test case] are satisfied.} & S1, S10, S2 \\ \hdashline
        $\drsh$ Repeatability & TS, TC & \multicolumn{2}{p{0.632\textwidth}}{A [test suite $\mid$ test case] should produce the same results under fixed conditions.} & S1, S9, S5 \\
        \hline
        
        \rowcolor[HTML]{ECF4FF} 
        \textbf{Resource efficiency} \tnote{*} & TS, TC & \multicolumn{2}{p{0.632\textwidth}}{The amount of resource (e.g., man-effort, computation, or test environment) needed by a [test suite $\mid$ test case] to execute.} & S1, S9, S5 \\ 
        \hline
        
        \rowcolor[HTML]{ECF4FF}
        \textbf{Reusability} \tnote{*} & TS, TC & The degree to which test cases of a test suite can be used as parts of another test suite. & The degree to which parts of a test case can be used as parts of another test case. & S1, S2 \\ \hdashline
        $\drsh$ Changeability & TS, TC & \multicolumn{2}{p{0.632\textwidth}}{The degree to which the structure and style of a [test suite $\mid$ test case] allow changes to be made easily, completely, and consistently.} & S1 \\ \hdashline
        $\drsh$ Universal & TS, TC & \multicolumn{2}{p{0.632\textwidth}}{The degree to which a [test suite $\mid$ test case] can be used in different test environments.} & S1 \\ 
        \hline
        
        \rowcolor[HTML]{ECF4FF}
        \textbf{Simplicity} \tnote{*} & TS, TC & How simple a test suite is in terms of the number of contained test cases, and execution steps. & How simple a test case is in terms of the test logic, the number of contained test cases, the number of steps. & S9, S10, S5, S11 \\ \hdashline
        $\drsh$ Single responsibility & TC & N/A & A test case should focus on verifying just one requirement, condition or behavior of SUT. & S10 \\ 
        \hline
        
        \rowcolor[HTML]{ECF4FF}
        \textbf{Usability} \tnote{*} & TS, TC & \multicolumn{2}{p{0.632\textwidth}}{How easy it is to execute a [test suite $\mid$ test case].}& S2 \\ \hdashline
        $\drsh$ Completeness & TS, TC & \multicolumn{2}{p{0.632\textwidth}}{A [test suite $\mid$ test case] should contain all relevant information for its execution.} & S1, S9, S5, S12 \\ \hdashline
        $\drsh$ Flexibility & TC & N/A & The degree of freedom in test execution that the test engineer can utilize. & S9, S2 \\ \hdashline
        $\drsh$ Learnability & TS, TC & \multicolumn{2}{p{0.632\textwidth}}{How easy it is to learn how to execute a [test suite $\mid$ test case].} & S1, S2 \\ \hdashline
        $\drsh$ Readability & TS, TC & \multicolumn{2}{p{0.632\textwidth}}{How expressive a [test suite $\mid$ test case] can be (using magic numbers, branching, and inexpressive naming conventions in test code can reduce its readability).} & S10, S13 \\ \hdashline
        $\drsh$ Understandability & TS, TC & \multicolumn{2}{p{0.632\textwidth}}{How easy it is for test users to know whether a [test suite $\mid$ test case] is suitable for their needs.} & S9, S2 \\ \hline   
    \end{longtable}
    \end{ThreePartTable}
}
\newpage
In the third section (Q8--9), we presented each respondent with the list of sub-quality attributes of the \textit{important} main attributes of either test cases (Q8) or test suites (Q9). 
For each corresponding sub-quality attribute, the respondents were asked to (i) select challenges that apply to the sub-quality attribute and (ii) describe any current support they have had to address the challenges.
At this point, free text fields were for respondents to describe the current support.
Meanwhile, the three challenges that respondents could choose from are: Challenge $C_{def}$ - \textit{Inadequate definition}, Challenge $C_{metric}$ - \textit{Lack of useful metrics}, and Challenge $C_{review}$ - \textit{Lack of an established review process}~\citep{juhnke2021challenges}.
Note that in the study of Juhnke et al., these challenges are listed as the types of challenges related to the quality assurance of test case specification.
Even though their context was test specifications, we decided to use them in our survey as they still cover three fundamental aspects: how to define the quality (in terms of quality attributes), how to measure the quality, and how to maintain such quality.
Nevertheless, the respondents could share extra information or specify different challenges in a free text field.

Note that Juhnke et al. also included concrete challenges under each of these types.
For example, the authors listed "Requirement coverage is the only known quality metric" as a concrete challenge under "Lack of useful metrics".
Here, we did not make use of the concrete challenges under each of these types as we did not want to limit the survey participants to fined-grain challenges that might not apply to their contexts while distracting them from seeing the whole picture.
Additionally, the respondents could select a ``None'' option in case they believed the quality attribute was not challenging for them to deal with.
Finally, the questionnaire concluded with closing questions for respondents to provide any extra comments (Q10) and contact information (Q11).
An example of a survey answer can be viewed via \url{https://doi.org/10.6084/m9.figshare.23309702.v1}.

\subsection{Survey instrument evaluation}\label{sec:surveyEvaluation}
We developed and improved the questionnaire based on discussions among the co-authors, covering the question logic, wording, and understandability.
Once the questionnaire was in a presentable state, we conducted a pilot study with seven external assessors (three researchers and four practitioners).
The questionnaire was refined based on their feedback, which was mainly about the clarity of terms and questions in the survey.

\subsection{Survey distribution}\label{sec:surveyDistribution}
The target population for the survey was software engineering professionals with working experience related to software testing.
They are developers, testers, test architects, etc.
To recruit participants, we followed suggestions given by Mello et al.~\citeyearpar{mello2015investigating}, who proposed a systematic approach to recruit participants for Software Engineering surveys from LinkedIn\footnote{\url{https://https://www.linkedin.com/}}.
Their proposed strategy helped us draw a large-scale sample and increase subject heterogeneity.

First, we used the Group Search feature in LinkedIn to search for groups related to software testing.
The search was conducted in July 2022 and returned 243 groups.
The first author read the groups' descriptions and excluded 40 groups based on the following exclusion criteria:
\begin{itemize}
    \item has a vague or no description;
    \item has its description out of the scope of software testing;
    \item restricted members to a region or country only;
    \item not in English;
    \item focuses on job advertisement or headhunting;
    \item targets researchers, students, or learners.
\end{itemize}

LinkedIn limits the maximum number of groups a LinkedIn user can be part of to 100 groups and the total number of pending requests to join to 20 groups only.
Hence, we sorted the 203 (243-40) remaining groups according to the number of members and then applied to join groups with the highest numbers of members first.
Our reasoning is that the higher the number of members in a group, the wider the coverage of relevant professionals in the group.
After three months of sending requests to join the sorted groups, the first author became a member of 20 groups (details in Table~\ref{tab:LinkedInMembersExtraction}).

{
    \footnotesize
    \begin{longtable}{p{0.01\textwidth}p{0.35\textwidth}p{0.1\textwidth}p{0.1\textwidth}p{0.1\textwidth}p{0.13\textwidth}}
    \caption{LinkedIn groups and members for the survey recruitment}
    \label{tab:LinkedInMembersExtraction}\\
    \hline
    \rowcolor[HTML]{D5D4D4} 
    \textbf{ID} & \textbf{Group name} & \textbf{Num of members} & \textbf{Sample size (in theory)} & \textbf{Sample size (in practice)} & \textbf{Distinct members to extract} \\ \toprule
    \endfirsthead
    \multicolumn{6}{c}%
    {Table \thetable\ continued from previous page} \\ \\
    \hline
    \rowcolor[HTML]{D5D4D4} 
    \textbf{ID} & \textbf{Group name} & \textbf{Num of members} & \textbf{Sample size (in theory)} & \textbf{Sample size (in practice)} & \textbf{Distinct members to extract} \\ \toprule
    \endhead
    \midrule
    \endfoot
    \bottomrule
    \endlastfoot
    \rowcolor[HTML]{EDEDED} 
    1 & \href{https://www.linkedin.com/groups/55636/}{Software Testing \& Automation} & 287,255 & 48 & 2,500 & 2,500 \\
    2 & \href{https://www.linkedin.com/groups/37631/}{Agile and Lean   Software Development} & 192,649 & 32 & 1,700 & 2,500 \\
    \rowcolor[HTML]{EDEDED} 
    3 & \href{https://www.linkedin.com/groups/60879/}{Quality Assurance -- QA Professional, Testing, Test Automation} & 129,788 & 22 & 1,149 & 1,500 \\
    4 & \href{https://www.linkedin.com/groups/23402/}{Software Testing and Quality Assurance group} & 71,113 & 12 & 634 & 1,000 \\
    \rowcolor[HTML]{EDEDED} 
    5 & \href{https://www.linkedin.com/groups/25412/}{Ministry of Testing -- the online software testing community} & 48163 & 8 & 432 & 1,000 \\
    6 & \href{https://www.linkedin.com/groups/2102114/}{Selenium Testing} & 36,521 & 6 & 330 & 1,000 \\
    \rowcolor[HTML]{EDEDED} 
    7 & \href{https://www.linkedin.com/groups/112164/}{Software Testing and QA} & 28,925 & 5 & 254 & 700 \\
    8 & \href{https://www.linkedin.com/groups/81915/}{Software Testing   Profession} & 24,114 & 4 & 222 & 700 \\
    \rowcolor[HTML]{EDEDED} 
    9 & \href{https://www.linkedin.com/groups/69745/}{QA/Testing} & 23,343 & 4 & 215 & 700 \\
    10 & \href{https://www.linkedin.com/groups/2647595/}{Quality Assurance Testing} & 16,060 & 3 & 143 & 500 \\
    \rowcolor[HTML]{EDEDED} 
    11 & \href{https://www.linkedin.com/groups/1421517/}{UK \& Europe Software Test \& QA Forum} & 13,346 & 2 & 117 & 500 \\
    12 & \href{https://www.linkedin.com/groups/2992544/}{Women In Software Engineering (WISE)} & 11,592 & 2 & 102 & 500 \\
    \rowcolor[HTML]{EDEDED} 
    13 & \href{https://www.linkedin.com/groups/6702200/}{Automation \& Manual Testing Group} & 6,534 & 1 & 57 & 100 \\
    14 & \href{https://www.linkedin.com/groups/2513515/}{Junior Testers} & 4,257 & 1 & 37 & 100 \\
    \rowcolor[HTML]{EDEDED} 
    15 & \href{https://www.linkedin.com/groups/2126747/}{Software Test Engineering} & 3,111 & 1 & 27 & 100 \\
    16 & \href{https://www.linkedin.com/groups/2921092/}{Test People - Testing Professional Open Networkers} & 2,787 & 0 & 24 & 100 \\
    \rowcolor[HTML]{EDEDED} 
    17 & \href{https://www.linkedin.com/groups/9081506/}{Software Testing Bootcamp} & 2,525 & 0 & 22 & 100 \\
    18 & \href{https://www.linkedin.com/groups/2674/}{Association for   Software Testing} & 1,551 & 0 & 14 & 100 \\
    \rowcolor[HTML]{EDEDED} 
    19 & \href{https://www.linkedin.com/groups/101675/}{Modern QA \& Testing} & 1,298 & 0 & 11 & 100 \\
    20 & \href{https://www.linkedin.com/groups/9010833/}{Software Development Engineer in Test $\parallel$ Test Automation $\parallel$ Quality Assurance Tester/Analyst} & 1,051 & 0 & 9 & 100 \\
    \rowcolor[HTML]{EDEDED} 
     & Total & 905,983 & 151 & 8,000 & 13,900
    \end{longtable}
}

Considering the gross number of members (905,983) of the 20 accepted groups already covers 81,53\% of the gross number of the 203 groups (1,111,218) and the time constraint of the survey study, we believed that recruiting participants from the 20 groups would be sufficient; hence, we stopped sending requests to join the other groups.

Having the 20 groups (details in Table~\ref{tab:LinkedInMembersExtraction}) as our sampling frame, the gross number of members of the 20 groups was 905,983 (accessed on 2022-07-15).
Similar to Mello et al.~\citeyearpar{mello2015investigating}, we used Bartlett et al.'s formula~\citep{kotrlik2001organizational} to calculate our theoretical sample size.
We aimed for a confidence level of 95\%, a confidence interval of 8, and the percentage occurrence of a state or condition of 0.5 (recommended by Bartlett et al.), and hence, our theoretical sample size was 151.
To avoid selection bias, for each selected group \textit{i}, the number of distinct members $\mathbf{X_i}$ to be added to the sample was in proportion with the gross number of members in that group, that is:

\begin{align*}
\footnotesize
    \mathbf{X_i} = \frac{\mathbf{GS_i}}{\mathbf{GS}} \times \mathbf{SS}
\end{align*}
\noindent where
\begin{tabular}{lp{0.7\textwidth}}
\footnotesize
\textit{$\mathbf{X_i}$} & is the number of distinct members in group \textit{i} to be added to the sample\\
\textit{$\mathbf{GS_i}$} & is the gross number of group \textit{i}\\
\textit{$\mathbf{GS}$} & is the total gross number of members of all 20 groups.\\
\textit{$\mathbf{SS}$} & is the theoretical optimum sample size (151).\\
\end{tabular}

Table~\ref{tab:LinkedInMembersExtraction} shows the number of distinct members from each selected group added to the sample size (151).
Note that a larger sample size gives a lower margin of error and higher confidence.
However, it also requires a larger recruitment size, which was, in our case, not feasible considering the time required to send LinkedIn messages manually.

Our next step was to extract distinct members from each selected group for recruitment (invitation).
According to Mello et al.~\citeyearpar{mello2015investigating}, the average completion rate through professional social networks lies between 3\% and 4\%.
Hence, by setting an expected completion rate of 3\%, our first estimation of the recruitment size was 5033 practitioners to achieve the stated sample size (151).
After sending the survey invitations to the first selected group, we noticed that while our response rate was around 9.2\%, our completion rate lay around 2\% instead of 3--4\% as estimated initially.
Hence, we increased our recruitment size to 8000 to make sure that we could acquire the planned sample size of 151.

In principle, we could have sent survey invitations to the first 8,000 practitioners extracted from the selected groups.
However, LinkedIn tends to have a bias in presenting group members~\citep{mello2015investigating} (i.e., tend to present members who connect directly to the user (first degree) or to other members who connect directly to the user (second degree)).
To mitigate this issue, we extracted more than 8,000 LinkedIn profiles so that we could randomly select the required number of distinct members~\citep{Baltes2022Sampling}.
In this recruitment process, before randomly selecting members, we first removed duplicate members and excluded members who had no description or whose descriptions did not indicate any working experience in software testing, i.e., recruiters and students.
The number of distinct members we recruited (invited) from each selected group is shown in the last column in Table~\ref{tab:LinkedInMembersExtraction}.

Besides using LinkedIn as a source for sampling, we also used personal networks and searched for relevant practitioners from the Internet to conduct convenience sampling and snowball sampling.
We used these non-probabilistic sampling approaches to increase the total number of responses.
It is also worth noting that we assigned a different survey link to each LinkedIn group and to each direct contact or source of contacts for traceability.

The data collection started on September 21, 2022 and ended on January 31, 2023. In total, we collected 828 responses, of which 183 were complete (meaning all the questions were answered).
Among these 828 responses, 35 were through personal networks, and 793 were through LinkedIn.

\subsection{Data analysis}\label{sec:dataAnalysis}
In total, we obtained 828 registered responses, of which 354 are valid, including 183 complete responses and 171 partial responses that completed at least Q6 or Q7 (the questions on which respondents were asked to rank the importance of the main quality attributes of either test cases (Q6) or test suites (Q7)).
We did not discard these 171 partial responses as they potentially contain information to answer RQ2.

For the two context dimensions, namely the size of SUT and the size of the test suite, we received only 31 and 25 responses, respectively, for the test suite data set, and 52 and 40 responses, respectively, for the test case data set.
Most respondents commented that it was not easy for them to know the sizes or they could not obtain such information.
Since the sizes of the data sets are too small to yield statistically significant results regarding the correlation between these two dimensions and the perceived importance, we decided not to investigate these two dimensions in this study.

Similarly, more than 90\% of the respondents did not provide details regarding any current support they have had to tackle the challenging quality attributes.
Since this question was asked at the end of the survey, we suspect that the respondents wanted to conclude the survey and, hence, did not answer this question.
As the provided information regarding the support is not enough for us to draw any significant conclusion, we also decided not to focus on this aspect in this study.

We started the data analysis by examining the ``Other'' answers, i.e., answers containing free texts, for the survey questions Q2--4 (eliciting context information) and Q8--9 (eliciting challenges).
Using a thematic coding approach~\citep{cruzes2011recommended} to analyze the free text answers, we studied each ``Other'' answer to see if we could consolidate the answer with an existing answer.
If this was not the case, i.e., the answer presented new information, we created a new code under the corresponding question.

Consequently, the free text answers to Q2--4 of 83\% responses (293 out of 354 valid responses) provided sufficient information to form reasonable software-testing contexts for the analysis to answer RQ2.
The remaining 17\% were incomplete or lacked contextual information to interpret the answers and were excluded from the analysis.
An example of excluded answers is ``Working across multiple teams as well for system testing.'' for the survey question Q2, which asks participants about their testing activities.
Another excluded one is ``No control of this'' as the answer to the type of development process for the system under test (part of question Q3.1).
Note that we did not consider these remaining 17\% as invalid responses as it could be that the respondent and us did not share the same understanding of the related concepts.
Hence, we still used these responses to answer the other research questions as long as the research questions did not require information regarding software-testing contexts.

With respect to the survey questions Q8--9, we did not receive any responses with only insufficient free text answers.
It means that all responses containing answers to Q8--9 were used for the analysis to answer RQ3.
Since RQ1 did not require answers (Q2--4) or (Q8--9), a response was included in the data set to answer RQ1 as long as it contained answers to Q6 or Q7.
As a result, we had 354 responses to answer RQ1, 293 responses to answer RQ2, and 183 responses to answer RQ3.

\begin{figure}
\begin{center}
\includegraphics[width=1\linewidth]{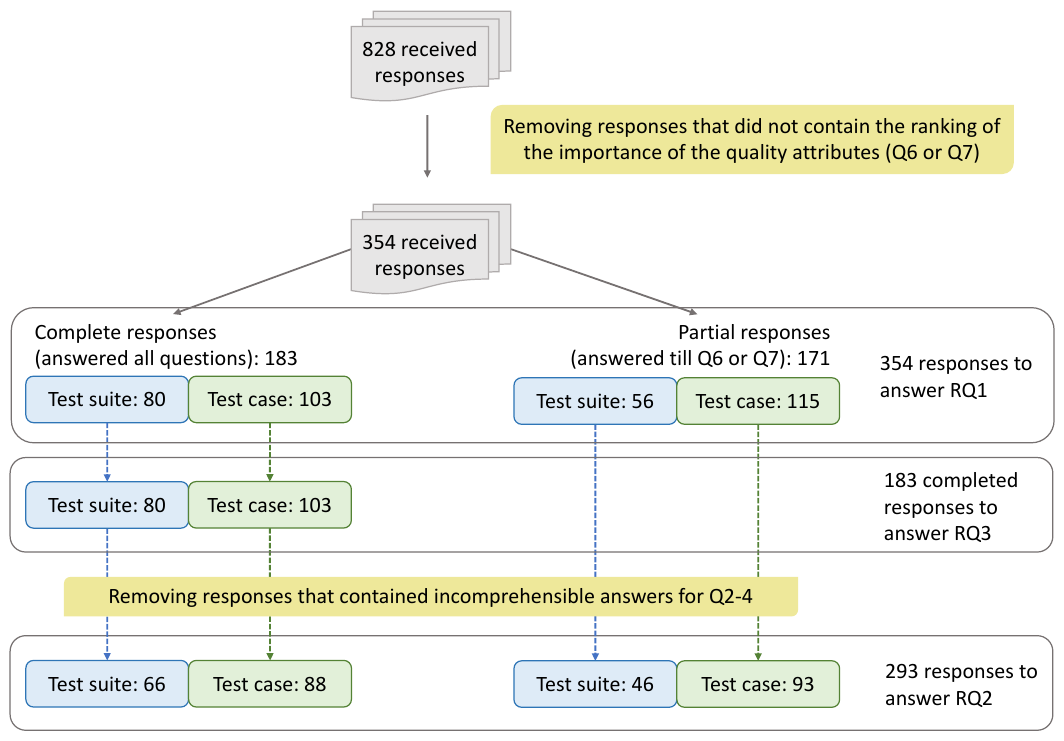}
\caption{Survey responses to answer research questions}
\label{fig:responseSummary}
\end{center}
\end{figure}

It is also worth noting that since our survey required the respondents to select either to answer questions regarding the quality of test cases or of test suites (Q5), the total responses actually contained two exclusive data sets: one data set regarding the test case quality and one data set regarding the test suite quality.
To ease the discussion, we call the two data sets the test case data set and the test suite data set.
Figure~\ref{fig:responseSummary} illustrates our responses filtering step with respect to the two mentioned data sets.
The two data sets are combined to answer RQ1.
With RQ2 and RQ3, we analyzed each data set independently.

\subsubsection{Statistical tests}
We followed Sheskin's guideline~\citep{sheskin2020handbook} to select statistical tests and then conducted the tests in IBM SPSS Statistics 28 to answer the research questions.
Generally, to decide which statistical tests to use, we assessed three points: (1) the study design (associations, predictions, group differences, reliability, etc.), (2) the type of collected data, and (3) the number of independent and dependent variables~\citep{sheskin2020handbook}.
Below, we explained our selection of tests and our reasoning for each research question.

\paragraph{RQ1}
First, we conducted a chi-square test of independence (at a 95\% level of significance) to check whether some quality attributes (of either test cases or test suites) were perceived as significantly more important than the other attributes.
Then, we performed post hoc comparisons of rates of importance votes by pairs of quality attributes.
It is worth noting that each post hoc comparison was a separate chi-square test on each pair of quality attributes.
Hence, we used Bonferroni correction to control Type I error across all of these post hoc tests.

Second, to verify if novice practitioners would be more likely to classify more quality attributes as important than senior practitioners, for each data set, we ran a Pearson's correlation~\citep{sheskin2020handbook} to assess the relationship between working experience and the percentage of quality attributes classified as important.

\paragraph{RQ2}
We first ran TwoStep Cluster analysis (provided by SPSS Statistics 28) on each data set (the test case data set and the test suite data set) to explore any possibility of grouping the respondents according to their ranking of the importance of the quality attributes.
This clustering step allowed us to observe similarities and differences in the software-testing contexts between the groups of respondents with different views on the importance of the quality attributes. 
To verify if such similarities and differences in the contexts could actually influence how important quality attributes were perceived, we conducted the Ordinal regression models and Multinominal Regression models~\citep{sheskin2020handbook}.
The models were constructed for the test case data set and the test suite data set separately.

The regression model~\citep{sheskin2020handbook} took into account all the context dimensions (multiple independent variables) and one quality attribute (one dependent variable).
Hence, there were nine regression models for nine quality attributes of test suites and seven models for seven quality attributes of test cases.
Note that we did not include the context dimension \textit{Testing tool} in the regression models.
It is because this dimension is dependent on the other context dimensions (the testing tool selection depends on the testing level, practice, etc.), and hence, violated one of the assumptions of the ordinal logistic regression test.

With each regression model, we first checked the result of the test of parallel lines to verify whether the assumption of proportional odds is met.
If the result of the test of parallel lines was \textbf{not} statistically significant ($p > 0.05$), meaning that the required assumption was met, we proceeded with checking the result of the Model Fitting Information, which is an overall measure of whether the model fits the data well.
If the significance value indicated in the Model Fitting Information was statistically significant ($p < 0.05$), we examined the results of the Tests of Model Effects, which report whether each independent variable has a statistically significant effect on the dependent variable ($p < 0.05$).
We used the results of the Tests of Model Effects to verify the correlation between the context dimensions and the quality attributes' importance levels.

If the required assumption of proportional odds was not met, i.e., the result of the test of parallel lines was statistically significant ($p < 0.05$), we instead ran multinomial logistic regression for the corresponding quality attribute.
Note that multinomial logistic regression does not consider the ordinal nature of our dependent variables (the importance levels of the quality attributes).
However, it is recommended as an alternative if the ordinal model does not meet the assumption of proportional odds~\citep{garcia2013statistical}.
For each multinomial regression model, we first checked the Model Fitting Information, which is an overall measure of whether the model fits the data well (same as for the Ordinal regression models).
If the significance value indicated in the Model Fitting Information was statistically significant ($p < 0.05$), we examined the results of the Likelihood Ratio Tests, which report whether each independent variable has a statistically significant effect on the dependent variable ($p < 0.05$).
The results of the Likelihood Ratio Tests helped us verify the correlation between software-testing contexts and the quality attributes' importance levels.

\paragraph{RQ3}
Statistically, we needed a test to determine whether there was a statistically significant difference in the challenges votes among the quality attributes.
Hence, the one-way repeated measures analysis of variance (ANOVA), which is also referred to as a within-subjects ANOVA, was the most suitable test~\citep{sheskin2020handbook}.
Our continuous, dependent variable was the percentage of respondents voting for each challenge for each quality attribute, while our within-subject factor (the independent variable) consisted of categorical levels, i.e., the challenges.
Before running the one-way repeated measures ANOVA on each of the two data sets (test cases and test suites), we used Shapiro-Wilk's test ($p > 0.05$) to check whether each data set had outliers and was normally distributed. 
If the normality assumption is not met~\citep{sheskin2020handbook}, we used the Friedman test, which is commonly used instead of ANOVA.

\section{Results and analysis}\label{sec:results_analysis}
In this section, we first describe the respondents' demographics (working experience and software-testing contexts), then present results and analysis according to our three research questions.
Note that the analysis centers around the two data sets, the test case data set and the test suite data set.
For clarity, the test case data set contains answers from respondents discussing test case quality, while the test suite data set contains answers from respondents discussing test suite quality (as explained earlier in Section~\ref{sec:dataAnalysis}).

\subsection{Demographics}\label{sec:results_analysis_Demographics}
\begin{figure}
\begin{center}
\includegraphics[width=0.8\textwidth]{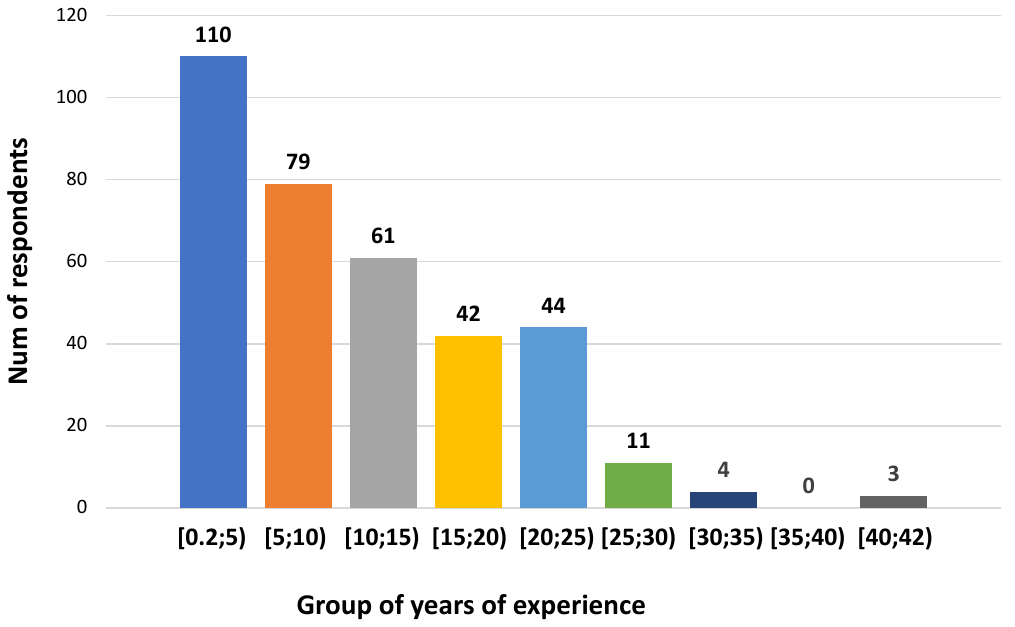}
\caption{Responses distribution over the working experience}
\label{fig:RQ0_Experience}
\end{center}
\end{figure}

Overall, the working experience of 354 respondents varied from less than one year to 42 years of experience, with a median of 8 years and a mean of 10.23 years; 90\% of the respondents had two to 25 years of experience.
Figure~\ref{fig:RQ0_Experience} illustrates the distribution of the respondents' working experience.
In this figure, we divide the respondents into bins (groups) with a uniform range of 5 years of experience.

\begin{figure}
\begin{center}
\includegraphics[width=1\textwidth]{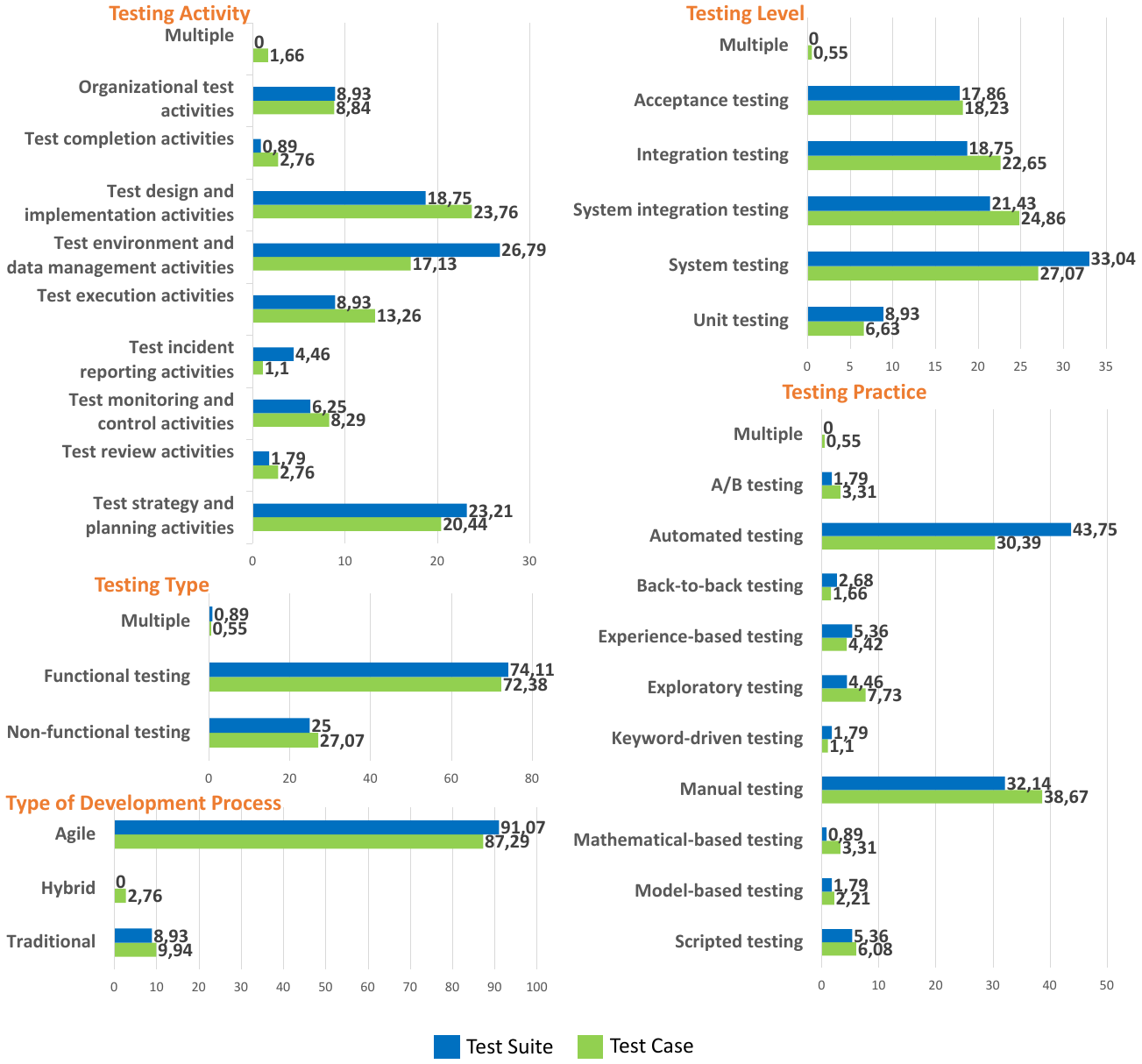}
\caption{Respondents' software-testing contexts in terms of Testing activity, Testing type, Type of development process, Testing level, and Testing practice}
\label{fig:RQ2_TS_TC_Context_1}
\end{center}
\end{figure}

\begin{figure}
\begin{center}
\includegraphics[width=1\textwidth]{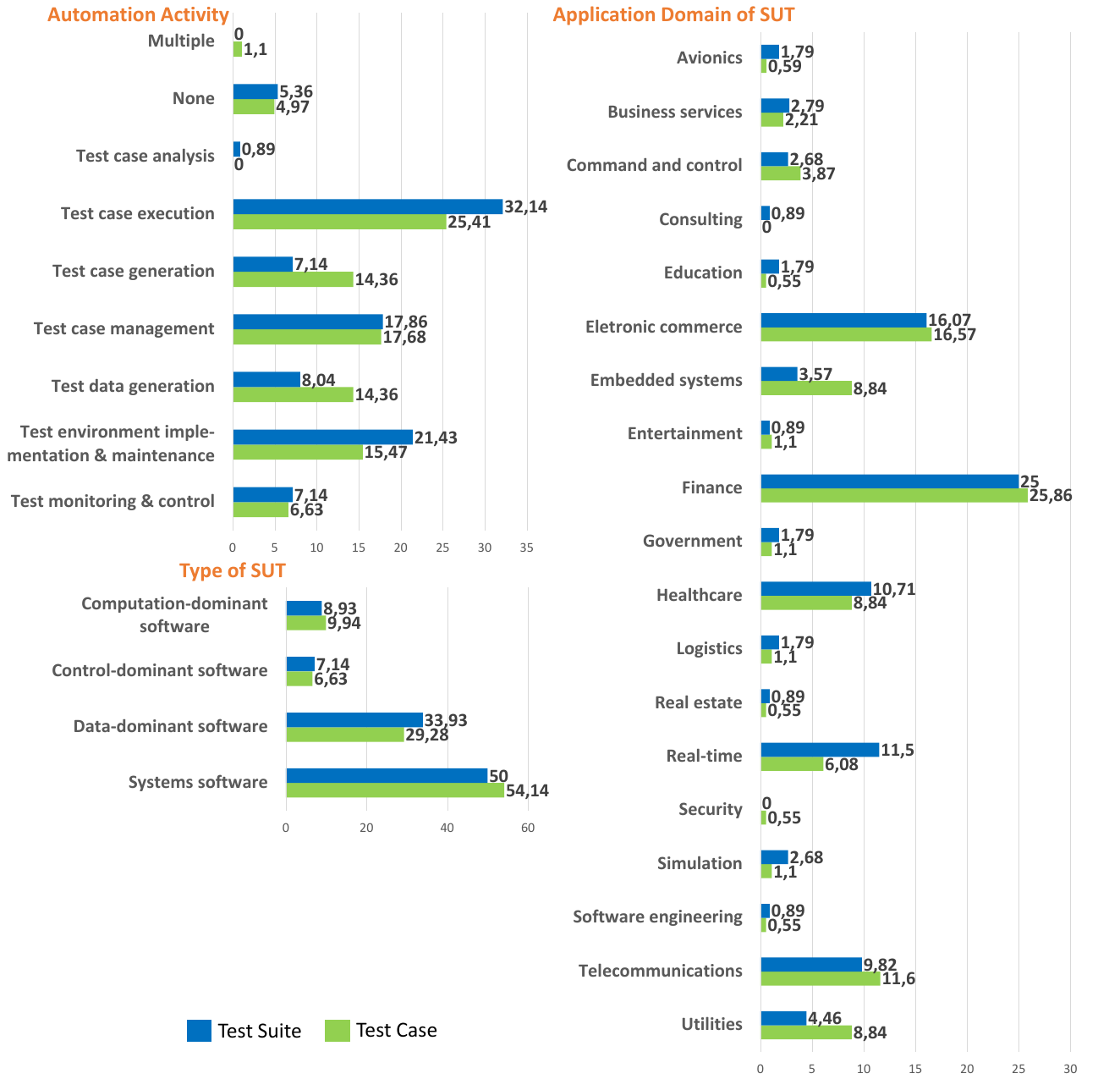}
\caption{Respondents' software-testing contexts in terms of Automation activity, Type of SUT, and Application domain of SUT}
\label{fig:RQ2_TS_TC_Context_2}
\end{center}
\end{figure}

Figure~\ref{fig:RQ2_TS_TC_Context_1} and Figure~\ref{fig:RQ2_TS_TC_Context_2} present the context information of the respondents under eight context dimensions (Testing activity, Testing level, Testing type, Testing practice, Automation activity, Application domain of SUT, Type of SUT, and Type of development process) for the test case data set and the test suite data set.

Our general observation, based on Figure~\ref{fig:RQ2_TS_TC_Context_1} and Figure~\ref{fig:RQ2_TS_TC_Context_2} is that the context information is highly similar from both data sets.
More specifically, the respondents from both data sets were mainly working at the \textit{system testing level} for \textit{functional testing}, applying \textit{automation in tests execution}.
They primarily tested \textit{systems software} in the \textit{finance} domain and used \textit{agile practices} in their development process.
The differences only come from their testing activities and testing practices.
The most common testing activities were \textit{test environment and data management activities} from the test suite data set and \textit{test design and implementation activities} from the test case data set.
For the testing practice, the respondents from the test suite data set did more automated testing, whereas those from the test case data set mainly did manual testing.

\subsection{RQ1 - Important quality attributes of test cases and test suites}\label{sec:results_analysis_RQ1}
With this research question, our focus was to investigate whether the importance of the quality attributes of test cases and test suites is perceived differently in practice.

\subsubsection{Differences in importance votes among quality attributes} 
\begin{figure}
\begin{center}
\includegraphics[width=1\linewidth]{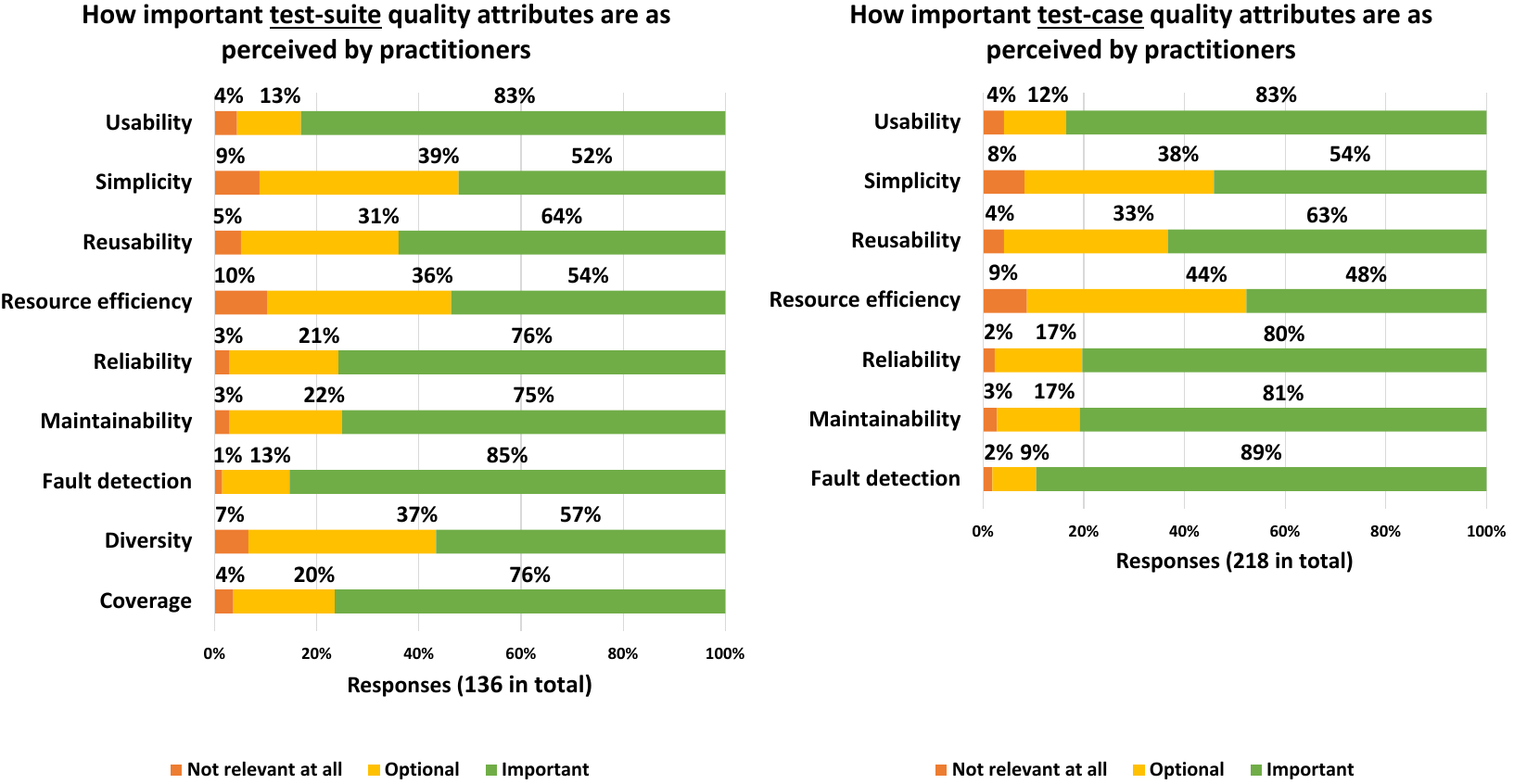}
\caption{Perceived importance of the quality attributes of test suites and test cases by practitioners}
\label{fig:RQ1_QAImportance}
\end{center}
\end{figure}

\hfill\\
We wanted to study if practitioners considered some quality attributes to be more important than others.
Hence, we focused on analyzing the similarities and differences in the rates of importance votes (under three importance levels: ``Important'', ``Optional'', and ``Not relevant at all'') among the quality attributes.
Since the quality attributes of test cases are different from those of test suites, we analyzed the two data sets: (1) the test case data set and (2) the test suite data set separately.
Figure~\ref{fig:RQ1_QAImportance} shows the percentage of votes (responses) each quality attribute of test cases and test suites receives under three importance levels: ``Important'', ``Optional'', and ``Not relevant at all''.

For test-suite quality, we can observe that the majority of respondents agreed that all test-suite quality attributes are important.
Based on Figure~\ref{fig:RQ1_QAImportance}, the quality attribute that the respondents regarded as the most important was Fault Detection (agreed by 85\% of the respondents), followed closely by Usability (83\%).
Likewise, Coverage, Reliability, and Maintainability were also regarded as important, as agreed by at least 75\% of the respondents.
On the other hand, Simplicity and Resource Efficiency were voted as the least important ones (52\% and 54\%, respectively).

We can also see that there were 15 respondents (out of 136) who marked at least two (out of nine) main quality attributes as ``Not relevant at all''.
Among these 15 respondents (ranging from two years to 25 years of experience), only two respondents (with twelve and 24 years of experience) considered six and five quality attributes as ``Not relevant at all'', three other respondents classified three attributes as ``Not relevant at all'', while the other ten marked two attributes as ``Not relevant at all''.
Based on this observation, we can see that the respondents who wanted to emphasize the importance of only a few quality attributes of test suites were from the minority group (2 respondents), while the rest of them regarded more attributes as important or at least optional.

We observe similar results for test-case quality.
While all quality attributes were perceived as important by many respondents, Fault Detection was regarded as the most important one (by 89\% of respondents), followed by Usability (83\%), Maintainability (81\%), and Reliability (80\%).
The data also shows Simplicity and Resource Efficiency were considered the least important attributes (54\% and 48\%, respectively).

Also, we note that 14 respondents (out of 218) classified at least two main quality attributes of test cases as ``Not relevant at all''.
Among these 14 respondents (ranging from 3 months to 25 years of experience), only two respondents (with 4 months and seven years of experience) considered six quality attributes as ``Not relevant at all'', five other respondents classified three quality attributes as ``Not relevant at all'', while the other seven marked two quality attributes as ``Not relevant at all''.
With this observation, similar to the case of test-suite quality, we can see that the respondents who wanted to emphasize the importance of only a few quality attributes of test cases were from the minority group (2 respondents), while the rest of them considered more attributes as optional or important.

The chi-square test of independence (at a 95\% level of significance) revealed statistically significant differences between seven quality attributes of \textit{test cases} regarding the importance votes (grouped by the three importance levels) ($\chi^{2}~(12)=170.504,~ p < 0.001 $).
Post hoc comparisons of rates of importance votes by pairs of \textit{test-case} quality attributes indicated that Fault Detection, Maintainability, Reliability, and Usability had significantly higher rates of votes for the highest importance level than Resource Efficiency, Reusability, and Simplicity (the test results can be viewed via \url{https://doi.org/10.6084/m9.figshare.23309702.v1}).

Likewise, the chi-square test of independence showed significant differences between nine quality attributes of \textit{test suites} regarding the importance votes ($\chi^{2}~(16)=87.282,~ p < 0.001 $).
Post hoc comparisons of rates of importance votes by pairs of \textit{test-suite} quality attributes revealed that Fault Detection, Usability, Coverage, Maintainability, and Reliability had significantly higher rates of votes for the highest importance level than Resource Efficiency and Simplicity (the test results can be viewed via \url{https://doi.org/10.6084/m9.figshare.23309702.v1}).
Additionally, Fault Detection and Usability had significantly higher rates of votes for the highest importance level than Diversity and Reusability.

\subsubsection{Perceived importance of quality attributes and practitioners' working experience}
\hfill\\
We wanted to study if novice practitioners would be more likely to classify more quality attributes as important than senior practitioners.
Since novice practitioners have less experience, and might have limited perspective regarding the quality of test cases and test suites, they might overestimate or underestimate the importance of quality attributes.
Senior practitioners, on the other hand, might want to focus more on a suitable selection of quality attributes for a better positive return on investment. 

Our interest in the association between the working experience and the selection of important quality attributes was inspired by studies~\citep{Chatzipetrou2020Component, Lo2015How} showing that practitioners' working experience could influence their perception of certain software-engineering-related aspects.
For example, Chatzipetrou et al.~\citeyearpar{Chatzipetrou2020Component} have shown that there exists a correlation between the working experience and the consideration of what software component attributes are important when selecting new components.

\begin{figure}
\begin{center}
\includegraphics[width=0.8\textwidth]{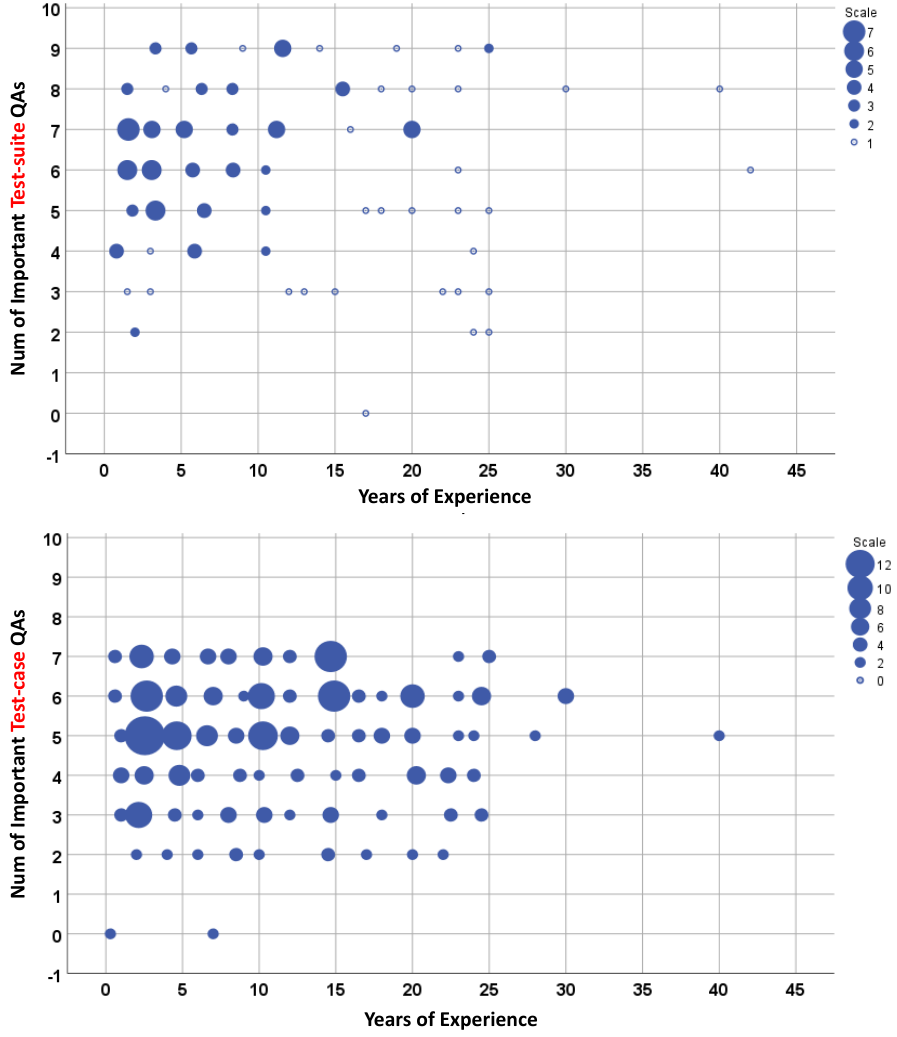}
\caption{Number of important quality attributes (QA) of test suites and test cases across years of experience}
\label{fig:RQ1_Experience_NumImportantQAs}
\end{center}
\end{figure}

Figure~\ref{fig:RQ1_Experience_NumImportantQAs} presents how the number of important quality attributes (of test case and test suite) changes with respect to the number of years of experience.
The dot scale illustrates how many respondents have the same number of years of experience while marking the same number of attributes as important.
For example, for test cases, the figure shows that one respondent with 40 years of experience marked five out of seven quality attributes as important, while twelve other respondents with two years and a half of experience marked five out of seven attributes as important.

Figure~\ref{fig:RQ1_Experience_NumImportantQAs} shows no clear pattern in how the number of important quality attributes varies in relation to the number of years of experience.
In other words, it is hard to tell if novice practitioners would consider more quality attributes as important than seniors or the other way around.
Hence, we ran a Pearson's correlation~\citep{sheskin2020handbook} for each data set (the test suite data set and test case data set) to assess the relationship between working experience and the percentage of quality attributes classified as important.


For each data set, the analyses showed the relationship to be linear with both variables normally distributed, as assessed by Shapiro-Wilk's test ($p > 0.05$), and there were no outliers.
The Pearson test indicated that there was no statistically significant correlation between working experience and the percentage of \textit{test-suite} quality attributes classified as important, $r(136) = -0.034, p = 0.698$.
Likewise, the Pearson test also showed that there was no statistically significant correlation between working experience and the percentage of \textit{test-case} quality attributes classified as important, $r(218) = -0.024, p = 0.723$.

\begin{center}
\fbox{\parbox{0.99\textwidth}{\textbf{Finding 1}:
With respect to test-case quality attributes, practitioners perceived Fault Detection, Maintainability, Reliability, and Usability as significantly more important than Resource Efficiency, Reusability, and Simplicity.
In the case of test-suite quality, Fault Detection, Usability, Coverage, Maintainability, and Reliability were significantly more important than Resource Efficiency and Simplicity.
Additionally, Fault Detection and Usability of test suites were significantly more important than Diversity and Reusability.
There was no statistically significant correlation between the working experience and the number of quality attributes classified as \textit{important}.}}
\end{center}

\subsection{RQ2 - Correlation between software-testing contexts and the perceived importance of quality attributes of test cases and test suites}\label{sec:results_analysis_RQ2}
With this research question, we focused on collecting and analyzing the relationship between the software-testing contexts of the respondents and the quality attribute classification outcomes.

First, for each data set, the \textit{TwoStep Cluster} analysis (provided by SPSS Statistics 28) returned two groups of respondents: the group of respondents who classified more quality attributes as important versus the other group of respondents.
The quality of the clustering result was considered fairly good (average silhouette = 0.2).

For the test-suite quality, to ease the discussion, we call the two groups returned by the TwoStep Cluster analysis Group TS\_1 (71 respondents) and Group TS\_2 (41 respondents).
While most of the respondents in Group TS\_1 agreed that all quality attributes of test suites should be important, most respondents in Group TS\_2 agreed that Diversity, Simplicity, Resource Efficiency, and Reusability should be optional or not relevant at all.

\begin{figure}
\begin{center}
\includegraphics[width=1\linewidth]{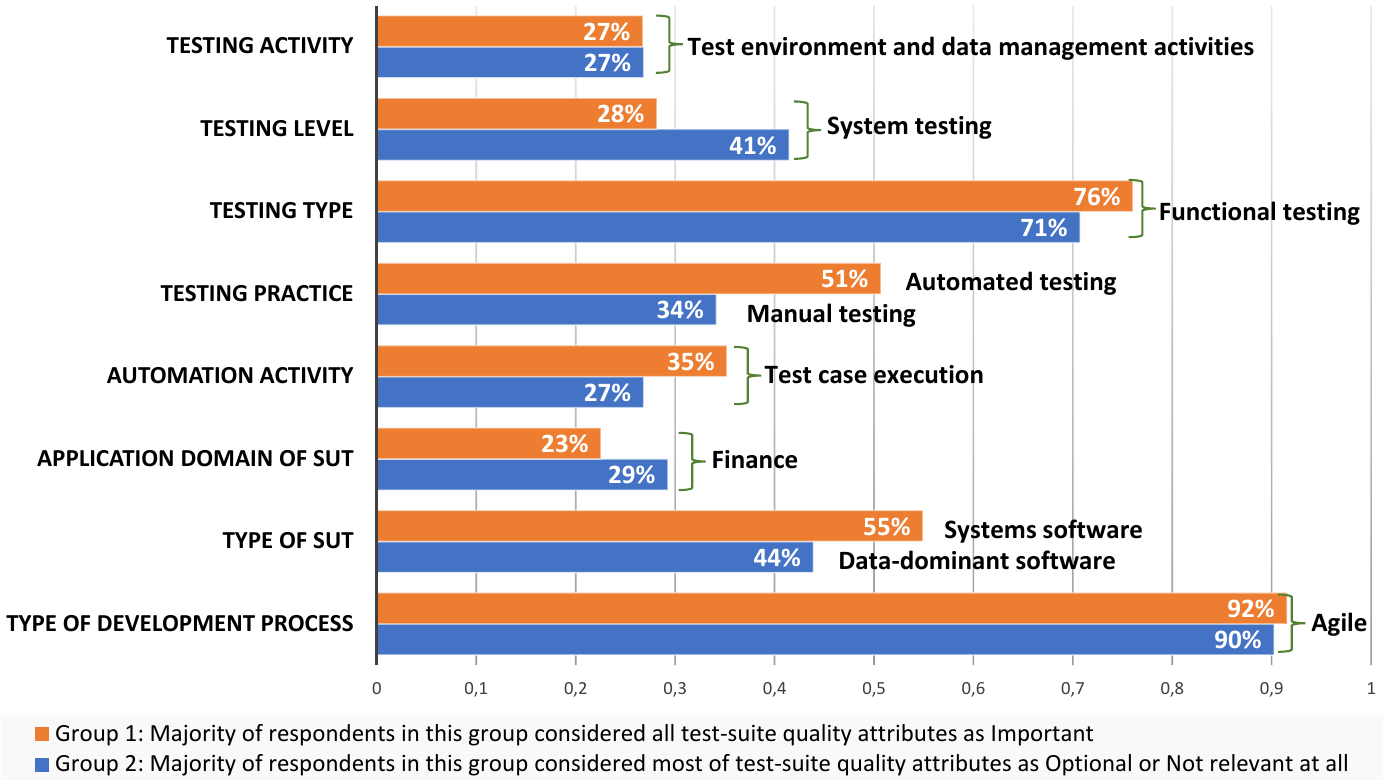}
\caption{Main differences in the software-testing contexts between respondents considering \textbf{test-suite} quality attributes (QAs) more important than those considering the QAs less important}
\label{fig:RQ2_TS_Context_QAImportance}
\end{center}
\end{figure}

Figure~\ref{fig:RQ2_TS_Context_QAImportance} illustrates the context differences between two groups TS\_1 and TS\_2.
We can see that the contexts between the two groups are quite similar and consistent with the common context reported earlier (detailed in Section~\ref{sec:results_analysis_Demographics}).
The main differences between these two groups are that the respondents from Group TS\_1 applied Automated testing as their testing practice while the respondents of Group TS\_2 focused more on Manual testing.
Moreover, Group TS\_1 mostly worked with Systems software and Group TS\_2 mainly worked with Data-dominant software.

For the test-case quality, also to ease the discussion, we call the two groups returned by the TwoStep Cluster analysis Group TC\_1 (71 respondents) and Group TC\_2 (110 respondents).
While most of the respondents in Group TC\_1 considered all quality attributes of test cases as important, respondents in Group TC\_2 agreed that Simplicity and Resource Efficiency should be optional or not relevant at all.

\begin{figure}
\begin{center}
\includegraphics[width=1\linewidth]{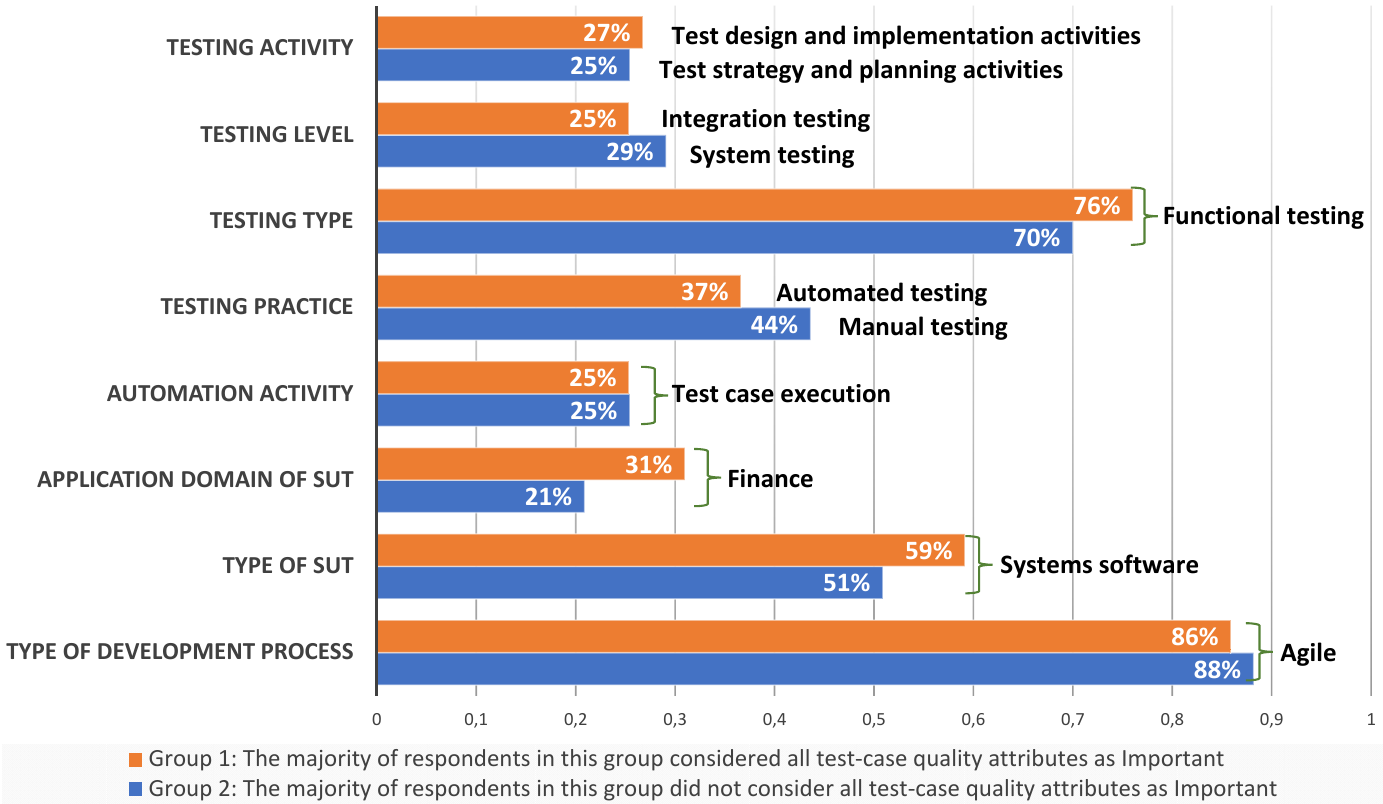}
\caption{Main differences in the software-testing contexts between respondents considering \textbf{test-case} quality attributes (QAs) more important than those considering the QAs less important}
\label{fig:RQ2_TC_Context_QAImportance}
\end{center}
\end{figure}

Figure~\ref{fig:RQ2_TC_Context_QAImportance} illustrates the context differences between two groups TC\_1 and TC\_2.
Even though the contexts between these two groups are rather similar, there are three context dimensions that distinguish the groups.
The three dimensions include Testing activity, Testing level, and Testing practice.
Most respondents from Group TC\_1 were involved in Test design and implementation activities, whereas respondents from Group TC\_2 mainly performed Test strategy and planning activities.
With the Testing level, the majority of Group TC\_1 focused on Integration testing, while most respondents from Group TC\_2 focused on System testing.
Furthermore, Automated testing was the main Testing practice of Group TC\_1, and Group TC\_2 mostly worked with Manual testing.

Overall, based on the outcome of the TwoStep Cluster analysis, despite the similarities, we can still see the differences in the software-testing contexts between the groups of respondents, which are distinguished by their diverse views on the importance of the quality attributes (of test cases and test suites).
To verify if such differences in the contexts could actually influence how important quality attributes were perceived, we conducted the Ordinal regression models and Multinominal Regression models~\citep{sheskin2020handbook}.
The models were constructed for the test case data set and the test suite data set separately.

Table~\ref{tab:RQ2_RegressionModel_Results} presents the results of the statistical tests.
For test-suite quality, the tests revealed that there were context dimensions that influenced the ranking of the importance of three quality attributes, namely Resource Efficiency, Reusability, and Usability.
More specifically, which Automation activity and Type of development process the respondents worked with had statistically significant effects on whether Resource Efficiency was perceived as important, optional, or not relevant at all.
Likewise, which Testing activity the respondents were involved with had a statistically significant effect on whether Reusability was perceived as important, optional, or not relevant at all.
In the case of test-suite Usability, even though the model met the assumption of proportional odds and the model fitted the data well, the convergence criteria were not satisfied.
Hence, no effect could be reported in this case.
Section~\ref{sec:discussion} discusses in detail how these context dimensions influence the perception of the quality attributes.

\begin{table*}[htbp]
  \centering
  \caption{Results of Ordinal regression model and Multinomial regression model to answer RQ2}  
    \resizebox{\columnwidth}{!}{
    \begin{tabular}{ll|l|l|r||r|r|}    
\cline{3-7}         
& & \multicolumn{3}{c||}{\cellcolor[rgb]{ .792,  .784,  .784}\textbf{Ordinal Regression Model}} & \multicolumn{2}{c|}{\cellcolor[rgb]{ .792,  .784,  .784}\textbf{Multinominal Regression Model}} \\\cline{3-7} 
& & \multicolumn{1}{p{8.89em}|}{\cellcolor[rgb]{ .851,  .851,  .851}\textcolor[rgb]{ .231,  .247,  .255}{\textbf{Step 1: Test of Parallel Lines}}} & \multicolumn{1}{p{10.39em}|}{\cellcolor[rgb]{ .851,  .851,  .851}\textcolor[rgb]{ .231,  .247,  .255}{\textbf{Step 2a: Model Fitting Information}}} & \multicolumn{1}{p{17.335em}||}{\cellcolor[rgb]{ .851,  .851,  .851}\textcolor[rgb]{ .231,  .247,  .255}{\textbf{Step 2b: Tests of Model Effects}}} & \multicolumn{1}{p{10.39em}|}{\cellcolor[rgb]{ .851,  .851,  .851}\textcolor[rgb]{ .231,  .247,  .255}{\textbf{Step 3a: Model Fitting Information}}} & \multicolumn{1}{p{17.335em}|}{\cellcolor[rgb]{ .851,  .851,  .851}\textcolor[rgb]{ .231,  .247,  .255}{\textbf{Step 3b: Likelihood Ratio Tests}}} \\
    \hline
     & \multicolumn{1}{|p{10em}|}{Coverage} & \multicolumn{1}{p{8.89em}|}{\cellcolor[rgb]{ .886,  .937,  .855}1.000} & \multicolumn{1}{p{10.39em}|}{\cellcolor[rgb]{ 1,  .949,  .8}0.175} &       &       &  \\
    \cline{2-7}
    & \multicolumn{1}{|p{10em}|}{Diversity} & \multicolumn{1}{p{8.89em}|}{\cellcolor[rgb]{ .886,  .937,  .855}0.548} & \multicolumn{1}{p{10.39em}|}{\cellcolor[rgb]{ .886,  .937,  .855}0.010} & \multicolumn{1}{p{17.335em}||}{\cellcolor[rgb]{ 1,  .949,  .8}$>$ 0.05} &       &  \\
    \cline{2-7}
    & \multicolumn{1}{|p{10em}|}{Fault detection} & \multicolumn{1}{p{8.89em}|}{\cellcolor[rgb]{ .886,  .937,  .855}1.00} & \multicolumn{1}{p{10.39em}|}{\cellcolor[rgb]{ 1,  .949,  .8}0.410} &       &       &  \\
    \cline{2-7}
    & \multicolumn{1}{|p{10em}|}{Maintainability} & \multicolumn{1}{p{8.89em}|}{\cellcolor[rgb]{ .886,  .937,  .855}0.988} & \multicolumn{1}{p{10.39em}|}{\cellcolor[rgb]{ 1,  .949,  .8}0.310} &       &       &  \\
    \cline{2-7}
    \parbox[t]{2mm}{\multirow{5}{*}{\rotatebox[origin=c]{90}{\textbf{Test-Suite QA}}}} & \multicolumn{1}{|p{10em}|}{Reliability (TS)} & \multicolumn{1}{p{8.89em}|}{\cellcolor[rgb]{ 1,  .949,  .8}$<$ 0.001} &       &       & \multicolumn{1}{l|}{\cellcolor[rgb]{ 1,  .949,  .8} 1.000} &  \\
    \cline{2-7}
      & \multicolumn{1}{|p{10em}|}{\cellcolor[rgb]{ .663,  .816,  .557}Resource efficiency} & \multicolumn{1}{p{8.89em}|}{\cellcolor[rgb]{ .886,  .937,  .855}0.381} & \multicolumn{1}{p{10.39em}|}{\cellcolor[rgb]{ .886,  .937,  .855}0.014} & \multicolumn{1}{p{17.335em}||}{\cellcolor[rgb]{ .663,  .816,  .557}Automation activity (p = 0.020)\newline{}Type of development process (p = 0.001)} & \cellcolor[rgb]{ 1,  1,  1} & \cellcolor[rgb]{ 1,  1,  1} \\
    \cline{2-7}
    & \multicolumn{1}{|p{10em}|}{\cellcolor[rgb]{ .663,  .816,  .557}Reusability} & \multicolumn{1}{p{8.89em}|}{\cellcolor[rgb]{ .886,  .937,  .855}0.882} & \multicolumn{1}{p{10.39em}|}{\cellcolor[rgb]{ .886,  .937,  .855}0.013} & \multicolumn{1}{p{17.335em}||}{\cellcolor[rgb]{ .663,  .816,  .557}Testing activity (p = 0.018)} & \cellcolor[rgb]{ 1,  1,  1} & \cellcolor[rgb]{ 1,  1,  1} \\
    \cline{2-7}
    &  \multicolumn{1}{|p{10em}|}{\cellcolor[rgb]{ .663,  .816,  .557}Simplicity} & \multicolumn{1}{p{8.89em}|}{\cellcolor[rgb]{ 1,  .949,  .8}$<$ 0.001} & \cellcolor[rgb]{ 1,  1,  1} & \cellcolor[rgb]{ 1,  1,  1} & \multicolumn{1}{p{10.39em}|}{\cellcolor[rgb]{ .886,  .937,  .855}$<$ 0.001} & \multicolumn{1}{p{17.335em}|}{\cellcolor[rgb]{ .663,  .816,  .557}Testing activity (p $<$ 0.001)\newline{}Application domain of SUT (p = 0.010)\newline{}Type of development process (p = 0.027)} \\
    \cline{2-7}
    & \multicolumn{1}{|p{10em}|}{\cellcolor[rgb]{ .663,  .816,  .557}Usability} & \multicolumn{1}{p{8.89em}|}{\cellcolor[rgb]{ .886,  .937,  .855}1.000} & \multicolumn{1}{p{10.39em}|}{\cellcolor[rgb]{ .886,  .937,  .855}0.001} & \multicolumn{1}{p{17.335em}||}{\cellcolor[rgb]{ 1,  .949,  .8}Some convergence criteria are not satisfied.} & \multicolumn{1}{l|}{\cellcolor[rgb]{ 1,  .949,  .8}0.391} & \cellcolor[rgb]{ 1,  1,  1} \\
    \hline
     & \multicolumn{1}{|l|}{Fault detection} & \cellcolor[rgb]{ 1,  .949,  .8}0.001 &       &       & \multicolumn{1}{l|}{\cellcolor[rgb]{ 1,  .949,  .8}0.577} &  \\
    \cline{2-7}
    & \multicolumn{1}{|l|}{Maintainability} & \cellcolor[rgb]{ .886,  .937,  .855}0.994 & \cellcolor[rgb]{ 1,  .949,  .8}0.097 &       &       &  \\
    \cline{2-7}
    \parbox[t]{2mm}{\multirow{5}{*}{\rotatebox[origin=c]{90}{\textbf{Test-Case QA}}}} & \multicolumn{1}{|l|}{Reliability} & \cellcolor[rgb]{ .886,  .937,  .855}0.999 & \cellcolor[rgb]{ 1,  .949,  .8}0.096 &       &       &  \\
    \cline{2-7}
    & \multicolumn{1}{|l|}{\cellcolor[rgb]{ .663,  .816,  .557}Resource efficiency} & \cellcolor[rgb]{ 1,  .949,  .8}0.003 & \cellcolor[rgb]{ 1,  1,  1} & \cellcolor[rgb]{ 1,  1,  1} & \multicolumn{1}{l|}{\cellcolor[rgb]{ .886,  .937,  .855}0.019} & \multicolumn{1}{p{17.335em}|}{\cellcolor[rgb]{ .663,  .816,  .557}Testing level (p = 0.003)\newline{}Testing practice (p = 0.008)\newline{}Automation activity (p $<$ 0.001)\newline{}Type of SUT (p = 0.026)} \\
    \cline{2-7}
    & \multicolumn{1}{|l|}{Reusability} & \cellcolor[rgb]{ .886,  .937,  .855}0.649 & \cellcolor[rgb]{ 1,  .949,  .8}0.071 &       &       &  \\
    \cline{2-7}
    & \multicolumn{1}{|l|}{Simplicity} & \cellcolor[rgb]{ .886,  .937,  .855}0.103 & \cellcolor[rgb]{ .886,  .937,  .855}0.014 & \multicolumn{1}{l||}{\cellcolor[rgb]{ 1,  .949,  .8}$>$ 0.05} &       & \\
    \cline{2-7}
    & \multicolumn{1}{|l|}{Usability} & \cellcolor[rgb]{ .886,  .937,  .855}0.996 & \cellcolor[rgb]{ 1,  .949,  .8}0.090 &       &       &  \\ \hline
    \end{tabular}
    }%
  \label{tab:RQ2_RegressionModel_Results}%
\end{table*}%

Regarding test-case quality, the four context dimensions, namely Testing level, Testing practice, Automation activity, and Type of SUT, had statistically significant effects on whether Resource Efficiency was perceived as important, optional, or not relevant at all.
No statistically significant result was found for the other quality attributes of test cases.

\begin{center}
\fbox{\parbox{0.99\textwidth}{\textbf{Finding 2}:
Software-testing contexts that practitioners were involved in did have an influence on their perception of the importance of certain quality attributes.
More specifically, the importance of Resource Efficiency, Reusability, Simplicity, and Usability of test suites depended on four context dimensions: Automation activity, Type of development process of SUT, Testing activity, and Application domain of SUT.
The importance of Resource Efficiency of test cases was affected by four context dimensions, namely Testing level, Testing practice, Automation activity, and Type of SUT.}}
\end{center}

\subsection{RQ3 - Challenges that apply to the important quality attributes of test cases and test suites}\label{sec:results_analysis_RQ3}
With this research question, we focused on collecting and analyzing the challenges that practitioners face in defining, measuring, and maintaining the important quality attributes.

Apart from the three provided challenges (Challenge $C_{def}$ - Inadequate definition, Challenge $C_{metric}$ - Lack of useful metrics, and Challenge $C_{review}$ - Lack of an established review process), we found that the other challenges given by the respondents (in free text) could be grouped together to form another generic challenge, that is $C_{support}$ - \textit{lack of external support} (with respect to a particular quality attribute).
This fourth generic challenge includes specific challenges given by respondents such as ``not enough time to achieve high maintainability'', ``lack of team members interested in achieving high quality'', ``lack of competence and experience of testing'', ``lack of product knowledge'', ``lack of support from team management'', ``Inadequate requirements definitions for new products'', etc.
Note that this fourth challenge does not cover the existing Challenge $C_{review}$ ``Lack of an established review process''.
Our three initially stated Challenges, together with the emerging one, represent four fundamental issues concerning a quality attribute of a test case or test suite, which are: cannot define the quality of test case or test suite quality (Challenge $C_{def}$), cannot measure the quality (Challenge $C_{metric}$), cannot review/maintain high quality (Challenge $C_{review}$), and cannot acquire support to achieve high quality (Challenge $C_{support}$).

\begin{figure}
\begin{center}
\includegraphics[width=\textwidth]{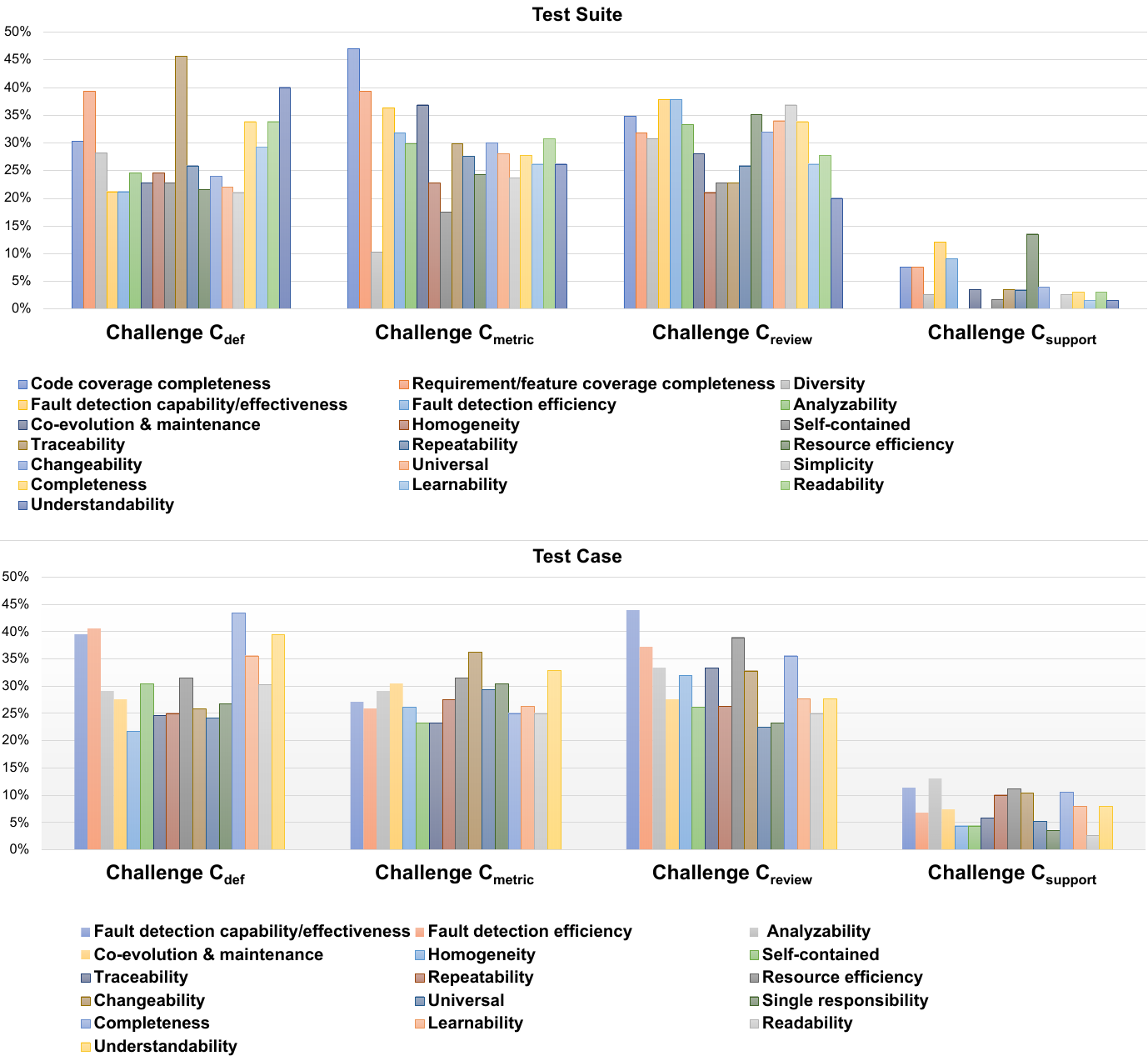}
\caption{Distribution of challenges that apply to the important quality attributes of test suites and test cases}
\label{fig:RQ3_QA_Challenges}
\end{center}
\end{figure}

Note that the respondents selected (or provided) challenges for the sub-quality attributes of each main quality attribute they marked as important.
Hence, to obtain an overview of how the four challenges are selected, for each sub-quality attribute, we calculated the percentage of respondents selecting each challenge.
The maximum number of votes that a sub-quality attribute received for a particular challenge was the number of respondents who classified the main quality attribute as important.
For example, Fault Detection (of test cases) was classified as important by 89 respondents, 35 out of which selected Challenge $C_{def}$ for Fault Detection Capability/Effectiveness.
Hence, we report that for Fault Detection Capability/Effectiveness, among all the four challenges, Challenge $C_{def}$ was voted by 39\% ($\mathbf{35\div89}$) of the corresponding respondents.
Figure~\ref{fig:RQ3_QA_Challenges} illustrates the calculation results.

As shown in Figure~\ref{fig:RQ3_QA_Challenges}, for both test cases and test suites, among Challenges $C_{def}$, $C_{metric}$, and $C_{review}$, none of the challenges received a remarkably higher percentage of voters than the others.
Compared to Challenges $C_{def}$, $C_{metric}$, and $C_{review}$, Challenge $C_{support}$, which emerged from the data collection, had a significantly lower percentage of voters.
Here, we were interested in verifying if this observation just happened by chance or if some challenge(s) were actually more common than others.
This question stemmed from the finding in the study of Juhnke et al.~\citeyearpar{juhnke2021challenges} where Challenges $C_{def}$, $C_{metric}$, and $C_{review}$ were originally reported.
Juhnke et al. found that Challenge $C_{review}$ appeared to be the most popular challenge, followed by Challenge $C_{metric}$, then Challenge $C_{def}$ in the context of test case specifications in automotive software testing.

Before running the statistical tests (ANOVA and Friedman tests) to verify our observation, we checked whether each data set (test cases and test suites) had outliers and was normally distributed. 
As a result, we found that the test suite data set had outliers and was not normally distributed, according to the assessed boxplot and Shapiro-Wilk's test ($p > 0.05$) respectively, while the test case data set met both assumptions, i.e., there were no outliers and the data was normally distributed.
Hence, we used the Friedman test, which is commonly used instead of ANOVA if the normality assumption is not met~\citep{sheskin2020handbook}, for the test suite data set and the one-way repeated measures ANOVA for the test case data set.

The results of the Friedman and ANOVA tests and their post-hoc tests are reported in Table~\ref{tab:RQ3_QA_Challenges_Friedman_ANOVA_Test}.
For the test case data set, Mauchly's test of sphericity indicated that the assumption of sphericity is violated ($p < 0.05$).
Therefore, we interpreted the results using the Greenhouse-Geisser correction.
Generally, we can see that there was a statistically significant difference in the challenge votes amongst the quality attributes.
In particular, the post-hoc analysis revealed statistically significant differences in the percentage of votes between Challenge $C_{support}$ and any other challenges (Challenges $C_{def}$, $C_{metric}$, and $C_{review}$) for both test cases and test suites.
The result is consistent with our data observation described previously.


\begin{table*}[htbp]
  \centering
  \caption{Results of one-way repeated measures ANOVA and Friedman test to answer RQ3}  
    \resizebox{\columnwidth}{!}{
    \begin{tabular}{|l|l|l|l|l|l|}
    \hline
    \multicolumn{3}{|c|}{\textbf{Test Suite}} & \multicolumn{3}{c|}{\textbf{Test Case}} \\
\cline{4-6}    \multicolumn{1}{|r}{} & \multicolumn{1}{r}{} &       & Mauchly's Test of Sphericity & \multicolumn{2}{l|}{$\chi^{2}~(5) = 14.90, p = .011$} \\
\cline{4-6}    \multicolumn{1}{|r}{} & \multicolumn{1}{r}{} &       & Greenhouse-Geisser & \multicolumn{2}{l|}{$\epsilon = 28.16$} \\
    \hline
    Friedman test & \multicolumn{2}{l|}{$\chi^{2}~(3) = 35.708, p < 0.001$} & ANOVA test & \multicolumn{2}{l|}{F(2.111, 31.667) = 101.822, p $<$ 0.001} \\
    \hline
    \multicolumn{1}{|l|}{\multirow{6}[2]{*}{Bonferroni post hoc test}} & Challenge $C_{def}$ - Challenge $C_{metric}$ & p = 1.000 & \multicolumn{1}{l|}{\multirow{6}[2]{*}{Bonferroni post hoc test}} & Challenge $C_{def}$ - Challenge $C_{metric}$ & p = 1.000 \\
          & Challenge $C_{def}$ - Challenge $C_{review}$ & p = 1.000 &       & Challenge $C_{def}$ - Challenge $C_{review}$ & p = 1.000 \\
          & Challenge $C_{metric}$ - Challenge $C_{review}$ & p = 1.000 &       & Challenge $C_{metric}$ - Challenge $C_{review}$ & p = 1.000 \\
          & \cellcolor[rgb]{ .886,  .937,  .855}Challenge $C_{support}$ - Challenge $C_{def}$ & \cellcolor[rgb]{ .886,  .937,  .855}p $<$ 0.0005 &       & \cellcolor[rgb]{ .886,  .937,  .855}Challenge $C_{support}$ - Challenge $C_{def}$ & \cellcolor[rgb]{ .886,  .937,  .855}p $<$ 0.001 \\
          & \cellcolor[rgb]{ .886,  .937,  .855}Challenge $C_{support}$ - Challenge $C_{metric}$ & \cellcolor[rgb]{ .886,  .937,  .855}p $<$ 0.0005 &       & \cellcolor[rgb]{ .886,  .937,  .855}Challenge $C_{support}$ - Challenge $C_{metric}$ & \cellcolor[rgb]{ .886,  .937,  .855}p $<$ 0.001 \\
          & \cellcolor[rgb]{ .886,  .937,  .855}Challenge $C_{support}$ - Challenge $C_{review}$ & \cellcolor[rgb]{ .886,  .937,  .855}p $<$ 0.0005 &       & \cellcolor[rgb]{ .886,  .937,  .855}Challenge $C_{support}$ - Challenge $C_{review}$ & \cellcolor[rgb]{ .886,  .937,  .855}p $<$ 0.001 \\
    \hline
    \end{tabular}%
    }
  \label{tab:RQ3_QA_Challenges_Friedman_ANOVA_Test}%
\end{table*}%

\begin{center}
\fbox{\parbox{0.99\textwidth}{\textbf{Finding 3}:
The three proposed challenges that apply to the important quality attributes of test cases and test suites (Challenge $C_{def}$ - Inadequate definition, Challenge $C_{metric}$ - Lack of useful metrics, Challenge $C_{review}$ - Lack of an established review process) were regarded as more or less equally by practitioners.
In addition, we identified an emerging challenge (Challenge $C_{support}$ - Lack of external support).}}
\end{center}

\section{Discussion}\label{sec:discussion}
We discuss in this section the two main aspects of our results, covering (i) the perceived importance of test-case quality attributes and test-suite quality attributes and their correlation with software-testing contexts, and (ii) the challenges that apply to the important quality attributes.

\subsection{The importance of quality attributes of test cases and test suites}
As reported in Section~\ref{sec:results_analysis_RQ1}, there was statistically significant evidence that Fault Detection, Maintainability, Reliability, and Usability were the most important quality attributes for both test cases and test suites. 
Likewise, Coverage, the quality attribute available for test suites only, was also one of the most important.
When investigating the hypothetical correlation between the importance of quality attributes and the context dimensions (Section~\ref{sec:results_analysis_RQ2}), we found that there were no statistically significant results on the influence of context dimensions on the perceived importance of these five important quality attributes (Fault Detection, Usability, Maintainability, Reliability, and Coverage).
One possible explanation is that for achieving high-quality test cases and test suites, some quality attributes should always be the main focus, independent of context.

On the other hand, other quality attributes, namely Diversity, Resource Efficiency, Reusability, and Simplicity, were the least important ones.
Furthermore, the difference in importance between the most and least important quality attributes was statistically significant.
Therefore, our findings suggest that not all quality attributes are regarded as equally important in practice.
This could be due to trade-offs between quality attributes.
This argument is in line with findings from several studies that have studied the trade-off between software quality attributes~\citep{sas2020Quality,wahler2012quality,feitosa2015investigating, ali2012testing}.
For example, to achieve high quality with respect to Fault Detection Capability and Efficiency, practitioners might need to utilize computation power and expand the test environment as much as needed, which in turn might reduce Resource Efficiency.
In another scenario, it could be that to improve Fault Detection Efficiency (``How quickly can a [test suite $\mid$ test cases] discover faults''), practitioners might need to combine several test cases with the same setup and teardown into one bigger test case.
In this second example, the Simplicity of test cases and test suites might reduce in proportion to the size of the merged test cases.

Besides the potential trade-off between quality attributes, the software-testing contexts can also be the reason why some attributes are considered more important than others (Section~\ref{sec:results_analysis_RQ2}).
Indeed, our statistical tests (Section~\ref{sec:results_analysis_RQ2}) showed that certain context dimensions could affect how practitioners perceived the importance of four quality attributes of test suites: Resource Efficiency, Reusability, Simplicity, and Usability as well as Resource Efficiency of test cases.
The following discussion touches on each of these quality attributes.

Particularly, regarding test-suite quality, the importance of Resource Efficiency significantly depended on what testing activity is automated and the type of development process of SUT.
Resource Efficiency is defined as ``The amount of resource (e.g., man-effort, computation, or test environment) needed by a test suite to execute''~\citep{tran2021assessing, tran2019test, adlemo2018test}.
Based on this definition, we can think of the following example.
In the context of automated test case execution, Resource Efficiency might be perceived as less important than in the context where test cases are generated automatically but have to be executed manually by practitioners.
Indeed, a practitioner explicitly mentioned that he or she ``works only with automatic tests maintained on the local server, and even if a test is not efficient (e.g., takes too long), it is acceptable. I might decide to run it, for example, once a week.''
Regarding the influence of types of the development process, if the SUT is developed in a traditional way (Waterfall, Structured systems analysis and design method, or no formal process) or in a hybrid way (a mix of traditional approaches and Agile practices), Resource Efficiency might be perceived as more important than in the context where SUT development followed Agile practices.
Generally, with the traditional or hybrid development approach, testing typically starts towards the end of the development life cycle after the coding has been done.
Hence, it is reasonable that in such a context, higher demand on resources is needed since there is more pressure to have the whole SUT tested adequately, i.e., Resource Efficiency becomes an important factor.

Likewise, the importance of test-suite Simplicity depended on the type of testing activities, the application domains, and the type of development process of the SUT.
Simplicity is defined as ``How simple a \textit{test suite} is in terms of the number of contained test cases, execution steps''~\citep{tran2019test, bowes2017how, adlemo2018test, kochhar2019practitioners}.
Hence, from the point of view of practitioners who are responsible for test design and implementation activities or test execution activities, Simplicity might be a more important requirement than for those who are responsible for test environment and data management activities.
Regarding application domains, it is known that for some domains, such as finance, as mentioned by one practitioner, it is common to outsource the development of some software components or the entire software system to a third party while the integration and customization would be done internally.
Hence, test cases are initially developed to test user acceptance scenarios in order to cover what business users would consider sufficient.
In this context, according to the same practitioner, ``more complicated tests will usually find more defects, such as pairwise testing mixes systems and devices to find more defects.
This is the situation when complicated tests are ``better than simple ones.''
In other words, the simplicity of test suites, according to the definition, might not be an important factor to consider in this context.
Regarding the types of the development process, it is reasonable to believe that the Simplicity of test suites could be perceived as more important by practitioners who followed Agile practices than those who have developed SUT using the traditional approach.
Overall, Simplicity in terms of code and design is one of the main principles of Agile methodologies that is seen as one of the main differences between Agile and traditional approaches.

Meanwhile, the importance of the Reusability of test suites depended on what testing activity practitioners were involved with (similar to the case of Simplicity discussed above).
For clarity, test-suite Reusability is defined as ``the degree to which test cases of a test suite can be used as parts of another test suite''~\citep{tran2021assessing, zeiss2007applying}.
Again, based on this definition, it is fair to argue that Reusability is more important to practitioners who have been involved in test design and implementation activities than other activities, such as test execution activities or test incident reporting activities.

Regarding the quality of test cases, we found that the importance of Resource Efficiency varied depending on four context dimensions: Automation Activity, Testing level, Testing practice, and Type of SUT.
On the one hand, similar to test suites, Automation Activity was one of the influencing context dimensions.
On the other hand, we had no statistically significant evidence regarding the influence of the Type of development process on this quality attribute's perceived importance as found for test suites.
For clarity, Resource Efficiency is defined as ``the amount of resource (e.g., man-effort, computation, or test environment) needed by a test case to execute.''~\citep{tran2021assessing, tran2019test, adlemo2018test}, which is similar for test suites.
We note here that the definitions of the quality attribute are similar between test cases and test suites, but the correlation findings are different.
However, it is worth emphasizing that the differences in findings do not imply conflicts in conclusions.
It just means that we did not obtain statistically significant evidence for some correlations.

With regard to Testing level, it is common that the execution time of unit test cases is shorter than that of higher-level test cases (such as system or system integration testing levels).
Therefore, Resource Efficiency might not necessarily be the main focus for unit test cases as for test cases of higher testing levels.
Regarding the testing practices, we argue that evidence-based testing or exploratory testing generally requires more resources (in terms of human effort) for test execution than other types of testing practices such as model-based testing or scripted testing.
Hence, practitioners working with former testing practices might regard Resource Efficiency as more important than those with the latter testing practices.
Finally, types of SUT could influence how the testing environments are set up.
For example, with parallel and distributed systems, setup and maintenance of a test environment are major cost drivers.
Resource Efficiency (in terms of test environments) for this type of SUT could be more important than other types of SUT.

\subsection{Challenges that apply to the important quality attributes}
In Section~\ref{sec:results_analysis_RQ3}, we report four challenges and how they were voted differently across the quality attributes of test cases and test suites.
Based on the collected data (illustrated in Figure~\ref{fig:RQ3_QA_Challenges}, amongst Challenges $C_{def}$ (Inadequate definition), $C_{metric}$ (Lack of useful metrics), and $C_{review}$ (Lack of an established review process), we did not see any challenge which was more popular or common that the others.
This finding suggests that practitioners still have difficulties with three fundamental aspects of test-case and test-suite quality, which are how to define, measure, and maintain the quality.

Meanwhile, Figure~\ref{fig:RQ3_QA_Challenges} shows that Challenge $C_{support}$ (Lack of means to achieve high quality) was significantly less common than Challenges $C_{def}$, $C_{metric}$, and $C_{review}$.
The results of the Friedman and one-way repeated measures ANOVA tests also supported this latter observation.
Nevertheless, we must remark that Challenge $C_{support}$ was the emerging information collected from the ``Other(s)'' free text option.
It means that the respondents were not informed about Challenge $C_{support}$ like the other challenges when answering the survey.
This can explain why Challenge $C_{support}$ appeared to be less common than the proposed challenges.
Nevertheless, the emergence of Challenge $C_{support}$ should not be neglected as it might be another common challenge but has not yet received much attention from researchers.
In other words, we argue that Challenge $C_{support}$ should be more carefully investigated as it could potentially open new research gaps.

\section{Limitations and threats to validity}\label{sec:validityThreats}
In this section, we discuss the limitations of our sampling method then three types of validity threats for survey studies based on Kitchenham and Pfleeger's guideline for personal opinion surveys~\citep{kitchenham2008personal}.

\subsection{Limitations of the sampling method}
There are two potential biases with our sampling approach: the bias due to the source of recruitment and the bias due to non-respondents, i.e., people who do not want to participate in the survey.
These two biases might prevent us from acquiring survey results that are representative of the population being studied. 

The first bias stems from the fact that we recruited participants mainly from LinkedIn.
Even though we also sent survey invitations via personal contacts, the corresponding responses were too few to have any significant impact.
There are other forums or platforms whose members may have different characteristics and views on the survey topic that we did not reach out to.
Additionally, LinkedIn has a bias in displaying group members.
We mitigated the issue from LinkedIn by extracting more members than required so that we could randomly select the needed number of people (as discussed in Section~\ref{sec:surveyDistribution}).
Another issue is that we could invite only members from the 20 LinkedIn groups of which the first author was a member.
To cope with this issue, we purposely joined groups with the highest number of participants and excluded groups having restrictions on languages, locations, or topics (details in Section~\ref{sec:surveyDistribution}).

The second bias is the self-selection bias, which is caused by us not being able to collect answers from non-respondents.
That is normally because they have mild or no interest in the survey topic~\citep{duda2010fallacy}, and hence, are not willing to join the survey.
People who responded to our survey invitation, by contrast, were likely to be more interested in the survey topic.
The potential problem here is that people without a vested interest in the topic might have different (or even opposite) opinions than those having an interest in the topic.
Since we could only send different survey links to different LinkedIn groups, we could not identify who did not respond to our survey within each group.
Therefore, sending a reminder to the non-respondents was not possible in our case.

\subsection{Validity threats}
Regarding the survey validity, we refer to Kitchenham and Pfleeger's guideline for personal opinion surveys~\citep{kitchenham2008personal} to discuss the three types of validity below.

\subsubsection{Content validity} 
Content validity concerns how appropriate the survey instrument is.
The assessment is typically done by a group of experts with relevant knowledge. 
The group ideally includes subject domain experts and members of the target population.
In our case, the survey instrument was developed and reviewed through iterations of discussion among the co-authors who have acquired relevant knowledge of the subject.
The instrument was also reviewed by practitioners of the target group in the pilot study.

\subsubsection{Construct validity} 
Construct validity concerns how well the survey instrument measures what it is designed to measure.
The major threat to our study is whether the respondents shared the same understanding of the context dimensions and context values as we did.
For example, Back-to-back testing, and Keyword-driven testing were some of the context values that respondents could select to describe their testing practices.
We assumed that the practitioners and we share the same understanding of these context values.
Nevertheless, this assumption might not always hold.
As pointed out by some respondents, they were not confident that they understood the information the same way we did.
We were aware of this potential threat but had to consider the trade-off between the survey length and the clarification of the context information.

To mitigate this threat, we used only peer-reviewed taxonomies for providing the context dimensions and context values.
Furthermore, we provided free text fields so that the respondents could express their doubts or different opinions.
Nevertheless, some of our respondents mentioned that our survey was quite complex and took them quite a lot of time to complete. 
Thus, we consider that our mitigation strategy has not fully addressed this threat.

Another threat to the construct validity is that we required each respondent to select only one context value under each context dimension.
This restriction was to support the analysis of the potential correlation between the quality attributes and the context dimensions (RQ2).
In practice, it is likely that respondents work in different contexts in parallel.
Hence, the restriction requires the respondents to be mindful and consistent when answering the questions relating to the importance of the quality attributes and their challenges so that their answers correspond to the selected context only.
To assist the respondents in focusing on one single context, we constructed a single matrix question covering all context dimensions in a table format (a matrix question) instead of having a question for each context dimension.
With this question design, the respondents could always review all their context answers in one view.

\subsubsection{Criterion validity} 
Criterion validity is about whether the survey instrument could distinguish respondents from different groups.
In our case, we used the demographics questions and the context dimensions to cluster respondents into different groups.
The context dimensions were extracted from the results of our tertiary study on the same topic~\citep{tran2021assessing}.
Three context dimensions (testing tools, test suite size, and SUT size) were removed from the analysis, and hence, one could argue that the respondents have not been well distinguished.
Additionally, this grouping was done by self-assessments and opinions.
The two issues could potentially introduce bias as we did not triangulate these self-assessments in our survey.

\section{Conclusion and future work}\label{sec:conclusion}
The benefits of having high-quality test artifacts have been well-studied in the literature such as for improving the productivity of the development teams and for helping developers keep up with the fast development pace.
However, the quality of test cases and test suites is composed of a set of roughly 30 quality attributes, according to our recent tertiary study on this topic.
It is difficult to focus on all of the quality attributes or, worse, on some random attributes.
Hence, to further assist practitioners in achieving high-quality test cases and test suites, it is better to recognize which quality attribute is important to practitioners but challenging for them to define, measure, and maintain.
Nevertheless, there have been no studies that have answered this question yet.
We, therefore, address this research gap in this industrial survey study.

We synthesized responses from 354 practitioners who got involved in various testing activities with a wide range of working experience.
Our novel contribution is presented in the answers to the research questions below.

RQ1: Our findings show that, generally, most practitioners perceived the quality attributes of test cases and test suites as important rather than optional or not relevant at all.
In a more fine-grained view, we found that Fault Detection, Usability, Maintainability, Reliability, and Coverage were regarded as the most important quality attributes, while Resource Efficiency and Simplicity were viewed as the least important attributes.

RQ2: The second important finding from this survey study is that the software-testing contexts practitioners were involved in did have an influence on their perception of the importance of quality attributes.
More specifically, the importance of Resource Efficiency, Reusability, Simplicity, and Usability of test suites depended on several context dimensions namely Automation activity, Type of development process of SUT, Testing activity, and Application domain of SUT.
Likewise, the importance of Resource Efficiency of test cases was affected by four context dimensions, namely Testing level, Testing practice, Automation activity, and Type of SUT.

RQ3: Our survey showed that practitioners still faced difficulties in the three main quality aspects, i.e., quality-attribute definition, measurement, and maintenance.
Additionally, they also emphasized the difficult nature of achieving high-quality test cases and test suites.

For researchers, our findings point out where practitioners actually need further support, i.e., the quality attributes that are important but challenging for them to define, measure, and maintain.
Our findings also highlight the important role of software-testing contexts in defining and assessing the quality of test cases and test suites.
Therefore, future research should (i) focus on these important quality attributes and (ii) further explore contextual differences when studying test-case and test-suite quality as such quality is composed of different quality attributes whose relevance is very likely to vary depending on the contexts.
For practitioners, our study encourages self-reflection.
Hopefully, our study can motivate organizations to take a step back and consider providing more support to practitioners to achieve high-quality test cases and test suites.
Along the same lines, we recommend organizations explicitly consider analyzing their software-testing contexts before searching for tools or different kinds of supports for assessing, improving, or maintaining the quality of test cases and test suites, which ultimately helps boost their confidence in testing.

\bmhead{Acknowledgements}
This work has been supported by ELLIIT; the Strategic Research Area within IT and Mobile Communications, funded by the Swedish Government. The work has also been supported by a research grant for the GIST (reference number 20220235) and SERT project from the Knowledge Foundation in Sweden.

\section*{Declarations}
\begin{itemize}
\item Funding:
This work has been supported by ELLIIT; the Strategic Research Area within IT and Mobile Communications, funded by the Swedish Government. The work has also been supported by a research grant for the GIST (reference number 20220235) and SERT project from the Knowledge Foundation in Sweden.

\item Conflict of interest/Competing interests:
The authors have no competing interests to declare that are relevant to the content of this article.

\item Ethics approval and consent to participate:
Before deciding to answer the survey questionnaire, the participants were informed that the survey was anonymous such that their identities were not retained (stated in the introduction of the survey instrument).

\item Consent for publication:
Before deciding to answer the survey questionnaire, the participants were informed that only aggregated results would be published in a scientific publication (stated in the introduction of the survey instrument).

\item Data availability 
As we only asked for the survey participants' consent to publish aggregated results, the original responses to the survey cannot be published.

\item Materials availability
The materials used to conduct this study can be assessed via \url{https://doi.org/10.6084/m9.figshare.23309702.v1}.

\item Code availability:
Not applicable

\item Author contribution:
Huynh Khanh Vi Tran, Nauman bin Ali, Michael Unterkalmsteiner, and Jürgen Börstler contributed to the study's conception and design. Material preparation and data collection were performed by Huynh Khanh Vi Tran, and the other authors discussed and reviewed all the steps. Huynh Khanh Vi Tran also led the data analysis with support from Panagiota Chatzipetrou. The first draft of the manuscript was written by Huynh Khanh Vi Tran, and all authors commented on intermediate versions of the manuscript. All authors read and approved the final manuscript.
\end{itemize}

\setlength{\bibsep}{2pt}
\bibliography{bib}

\end{document}